\documentclass[longauth]{aa}  

\usepackage{graphicx}
\usepackage{txfonts}
\usepackage{xspace}
\usepackage{placeins}
\usepackage[sticky-per=false,
    separate-uncertainty,
    angle-symbol-over-decimal,
    ]{siunitx}
\newcommand{\lya}{Ly$\alpha$\xspace}
\newcommand{\halpha}{H$\alpha$\xspace}
\newcommand{\hbeta}{H$\beta$\xspace}
\newcommand{\hi}{H\,{\sc i}\xspace}

\newcommand{\nii}{N\,{\sc ii}\xspace}
\newcommand{\oiii}{[O\,{\sc iii}]\xspace}

\newcommand{\ciim}{[C\,{\sc ii}]\,\SI{158}{\micro\meter}\xspace}
\newcommand{\cii}{[C\,{\sc ii}]\xspace}

\newcommand{\heii}{He\,{\sc ii}\xspace}
\newcommand{\civ}{C\,{\sc iv}\xspace}
\newcommand{\sysname}{J1000+0234\xspace}

\newcommand{\singlet}[3]{[#1\,{\sc #2}]\,$\lambda #3$}
\newcommand{\permitted}[3]{#1\,{\sc #2}\,$\lambda #3$}
\newcommand{\doublet}[4]{[#1\,{\sc #2}]\,$\lambda\lambda #3,#4$}

\DeclareSIUnit\jansky{Jy}
\DeclareSIUnit\erg{erg}
\DeclareSIUnit\mag{mag}
\DeclareSIUnit\parsec{pc}
\DeclareSIUnit\arcsec{arcsec}
\DeclareSIUnit\arcsect{^{\prime\prime}}
\DeclareSIUnit\msun{M_\odot}
\DeclareSIUnit\lumsol{L_\odot}
\DeclareSIUnit\zsun{Z_\odot}
\DeclareSIUnit\year{yr}
\DeclareSIUnit\beam{beam}
\DeclareSIUnit\angstrom{\text {Å}}

\defcitealias{VeilleuxAndOsterbrock1987SpectralClassGalaxies}{VO87}
\defcitealias{Baldwin1981BPTpaper}{BPT}
\usepackage{xcolor}
\usepackage{multirow}
\usepackage{natbib}
\bibpunct{(}{)}{;}{a}{}{,} % to follow the A&A style

\usepackage[colorlinks=true,
linkcolor=blue, citecolor=blue, filecolor=blue, urlcolor=blue,
unicode=true]{hyperref}
\begin{document}

% \title{Extreme [O III] emission in a \lya blob} 
\title{A hidden active galactic nucleus powering bright \oiii nebulae in a protocluster at $z=4.5$ revealed by JWST} 

   %\subtitle{}

   \author{M. Solimano\inst{1}
     \and
     J. Gonz\'alez-L\'opez\inst{2, 3}
     \and
     M. Aravena\inst{1}
     \and
     B. Alcalde Pampliega\inst{1,46}
     \and
     R. J. Assef\inst{1}
     \and
     M. B\'ethermin\inst{4,5}
     \and
     M. Boquien\inst{6}
     \and
     S. Bovino\inst{7, 8, 9}
     \and
     C. M. Casey\inst{10, 11}
     \and
     P. Cassata\inst{12,13}
     \and
     E. da Cunha\inst{14}
     \and
     R. L. Davies\inst{15,16}
     \and
     I. De Looze\inst{17,18}
     \and
     X. Ding\inst{19}
     \and
     T. D\'iaz-Santos\inst{20,21}
     \and
     A. L. Faisst\inst{22}
     \and
     A. Ferrara\inst{23}
     \and
     D. B. Fisher\inst{15,16}
     \and
     N. M. F{\"o}rster-Schreiber\inst{30}
     \and
     S. Fujimoto\inst{10}
     \and
     M. Ginolfi\inst{24,9}
     \and
     C. Gruppioni\inst{25}
     \and
     L. Guaita\inst{26}
     \and
     N. Hathi\inst{27}
     \and
     R. Herrera-Camus\inst{7}
     \and
     E. Ibar\inst{28}
     \and
     H. Inami\inst{29}
     \and
     G. C. Jones\inst{30}
     \and
     A. M. Koekemoer\inst{27}
     \and
     L. L. Lee\inst{31}
     \and
     J. Li\inst{14}
     \and
     D. Liu\inst{32}
     \and
     Z. Liu\inst{19,33,34}
     \and
     J. Molina\inst{28}
     \and
     P. Ogle\inst{27}
     \and
     A. C. Posses\inst{1}
     \and
     F. Pozzi\inst{34}
     \and
     M. Rela{\~n}o\inst{36,37}
     \and
     D. A. Riechers\inst{38}
     \and
     M. Romano\inst{41,13,39}
     \and
     J. Spilker\inst{40}
     \and
     N. Sulzenauer\inst{41}
     \and
     K. Telikova\inst{1}
     \and
     L. Vallini\inst{25}
     \and
     K. Vasan G. C.\inst{42}
     \and
     S. Veilleux\inst{43}
     \and
     D. Vergani\inst{25}
     \and
     V. Villanueva\inst{7}
     \and
     W. Wang\inst{44}
     \and
     L. Yan\inst{45}
     \and
     G. Zamorani\inst{25}
   }

\institute{
  Instituto de Estudios Astrof\'isicos, Facultad de Ingenier\'ia y Ciencias, % 1
  Universidad Diego Portales, Av.  Ej\'ercito Libertador 441, Santiago, Chile [C\'odigo Postal 8370191] \email{manuel.solimano@mail.udp.cl}
  \and
  Instituto de Astrof\'isica, Facultad de F\'isica, % 2
  Pontiﬁcia Universidad Cat\'olica de Chile, Santiago 7820436, Chile
  \and
  Las Campanas Observatory, Carnegie Institution of Washington, % 3
  Ra\'ul Bitr\'an 1200, La Serena, Chile
  \and
  Universit\'e de Strasbourg, CNRS, Observatoire astronomique de Strasbourg, UMR 7550, %4
  67000 Strasbourg, France
  \and
  Aix Marseille Univ, CNRS, CNES, LAM, Marseille, France % 5
  \and
  Universit\'e C\^ote d'Azur, Observatoire de la C\^ote d'Azur, CNRS, Laboratoire Lagrange, 06000, Nice, France % 6
  \and
  Departamento de Astronom\'ia, Facultad Ciencias F\'isicas y Matem\'aticas, % 7
  Universidad de Concepci\'on, Av. Esteban Iturra s/n Barrio Universitario, %
  Casilla 160, Concepci\'on, Chile
  \and
  Chemistry Department, Sapienza University of Rome, P.le A. Moro, 00185 Rome, Italy % 8
  \and
  INAF, Osservatorio Astrofisico di Arcetri, Largo E. Fermi 5, I-50125, Firenze, Italy %9
  \and
  The University of Texas at Austin, 2515 Speedway Blvd Stop C1400, Austin, TX 78712, USA % 10
  \and
  Cosmic Dawn Center (DAWN), R{\aa}dmandsgade 64, 2200 K{\o}benhavn N, Denmark % 11
  \and
  Dipartimento di Fisica e Astronomia, Universit{\`a} di Padova, % 12
  Vicolo dell’Osservatorio, 3, 35122 Padova, Italy
  \and
  INAF Osservatorio Astronomico di Padova, vicolo dell’Osservatorio 5, %13
  35122 Padova, Italy
  \and
  International Centre for Radio Astronomy Research (ICRAR), %14
  The University of Western Australia, M468, 35 Stirling Highway,%
  Crawley, WA 6009, Australia
  \and
  Centre for Astrophysics and Supercomputing, Swinburne Univ. % 15
  of Technology, PO Box 218, Hawthorn, VIC, 3122, Australia
  \and
  ARC Centre of Excellence for All Sky Astrophysics in % 16
  3 Dimensions (ASTRO 3D), Australia
  \and
  Sterrenkundig Observatorium, Ghent University, Krijgslaan% 17
  281-S9, B-9000 Ghent, Belgium
  \and
  Department of Physics \& Astronomy, University College London, % 18
  Gower Street, London WC1E\,6BT, UK  
  \and
  Kavli Institute for the Physics and Mathematics of the Universe, % 19
  The University of Tokyo, Kashiwa, Japan 277-8583 (Kavli IPMU, WPI)
  \and
  Institute of Astrophysics, Foundation for Research and % 20
  Technology-Hellas (FORTH), Heraklion, 70013, Greece
  \and
  Chinese Academy of Sciences South America Center for % 21
  Astronomy (CASSACA), National Astronomical Observatories, %
  CAS, Beijing, 100101, PR China
  \and
  Caltech/IPAC, MS 314-6, 1200 E. California Blvd. Pasadena, CA 91125, USA % 22
  \and
  Scuola Normale Superiore, Piazza dei Cavalieri 7, % 23
  I-50126 Pisa, Italy
  \and
  Dipartimento di Fisica e Astronomia, Universit{\`a} di Firenze, %24
  via G. Sansone 1, 50019 Sesto Fiorentino, Firenze, Italy
  \and
  INAF - Osservatorio di Astrofisica e Scienza dello Spazio di Bologna, % 25
  via Gobetti 93/3, 40129 Bologna, Italy
  \and 
  Universidad Andr\'es Bello, Facultad de Ciencias Exactas, % 26
  Departamento de F\'isica, Instituto de Astrof\'isica, %
  Fernandez Concha 700, Las Condes, %
  Santiago RM, Chile
  \and
  Space Telescope Science Institute, 3700 San Martin Dr., Baltimore, MD 21218, USA % 27
  \and
  Instituto de F\'isica y Astronom\'ia, Universidad de Valpara\'iso, % 28
  Avda. Gran Breta{\~n}a 1111, Valpara\'iso, Chile
  \and
  Hiroshima Astrophysical Science Center, Hiroshima University, %29
  1-3-1 Kagamiyama, Higashi-Hiroshima, Hiroshima 739-8526, Japan
  \and
  Department of Physics, University of Oxford, Denys Wilkinson Building, % 30
  Keble Road, Oxford OX1 3RH, UK
  \and
  Max-Planck-Institut f{\"u}r extraterrestrische Physik, % 31
  Gie{\ss}enbachstra{\ss}e 1, 85748 Garching, Germany
  \and
  Purple Mountain Observatory, Chinese Academy of Sciences, % 32
  10 Yuanhua Road, Nanjing 210023, China
  \and
  Center for Data-Driven Discovery, Kavli IPMU (WPI), UTIAS, %33 
  The University of Tokyo, Kashiwa, Chiba 277-8583, Japan
  \and
  Department of Astronomy, School of Science, The University of Tokyo, %34
  7-3-1 Hongo, Bunkyo, Tokyo 113-0033, Japan
  \and
  Dipartimento di Fisica e Astronomia, Universit{\`a} di Bologna, %35
  via Gobetti 93/2, 40129, Bologna, Italy
  \and
  Dept. Fisica Teorica y del Cosmos, Universidad de Granada, % 36
  Granada, Spain
  \and
  Instituto Universitario Carlos I de F\'{i}sica Te\'{o}rica % 37
  y Computacional, Universidad de Granada, %
  E-18071 Granada, Spain
  \and
  I. Physikalisches Institut, Universität zu K\"oln, Z\"ulpicher % 38
  Strasse 77, 50937 K\"oln, Germany 
  \and
  National Centre for Nuclear Research, ul. Pasteura 7, 02-093 Warsaw, Poland %39
  \and
  Department of Physics and Astronomy and George P. and Cynthia Woods %40
  Mitchell Institute for Fundamental Physics and Astronomy, %
  Texas A\&M University, 4242 TAMU, College Station, TX 77843-4242, US
  \and
  Max-Planck-Institut f{\"u}r Radioastronomie, Auf dem H{\"u}gel 69, Bonn, D-53121, Germany % 41
  \and
  University of California, Davis, 1 Shields Ave., Davis, CA 95616, USA % 42
  \and
  Department of Astronomy and Joint Space-Science Institute, % 43
  University of Maryland, College Park, MD 20742, USA
  \and
  Astronomisches Rechen-Institut, Zentrum f{\"u}r Astronomie der Universit{\"a}t Heidelberg, % 44
  M{\"o}nchhofstr. 12-14, 69120 Heidelberg, Germany
  \and
  Caltech Optical Observatories, California Institute of Technology, Pasadena, CA 91125, USA % 45
  \and
  ESO Vitacura, Alonso de Córdova 3107,Vitacura, Casilla 19001, % 46
  Santiago de Chile, Chile
}
   \date{Received -; accepted -}

   \abstract{
%	   Galaxy protoclusters at $z>3$ are often linked to large-scale emission from ionized gas, usually traced by hydrogen Lyman-$\alpha$ (\lya) extending on scales of tens to hundreds of kpc, and signature of a wide variety of physical processes happening within protoclusters.
           Galaxy protoclusters are sites of rapid growth, with a high density of massive galaxies driving elevated rates of star formation and accretion onto supermassive black holes.
   Here, we present new JWST/NIRSpec IFU observations of the J1000+0234 group at $z=4.54$, a dense region of a protocluster hosting a massive, dusty star forming galaxy (DSFG).
   The new data reveal two extended, high-equivalent-width (EW$_0>\SI{1000}{\angstrom}$) \oiii nebulae that appear at both sides of the DSFG along its minor axis (namely O3-N and O3-S).
   On one hand, the spectrum of O3-N  shows a broad and blueshifted component with a full width at half
maximum (FWHM) of $\sim \SI{1300}{\kilo\meter\per\second}$ , suggesting an outflow origin.
   On the other hand, O3-S stretches over \SI{8.6}{\kilo\parsec}, and has a velocity gradient that spans \SI{800}{\kilo\meter\per\second}, but shows no evidence of a broad component. 
   However, both sources seem to be powered at least partially by an active galactic nucleus (AGN), so we classified them as extended emission-line regions (EELRs).
   The strongest evidence comes from the detection of the high-ionization \singlet{Ne}{v}{3427} line toward O3-N, which paired with the lack of hard X-rays implies an obscuring column density above the Compton-thick regime.
    The [Ne\,\textsc{v}] line is not detected in O3-S, but we measure a \permitted{He}{II}{4687}/\hbeta$=0.25$, which is well above the expectation for star formation.
   Despite the remarkable alignment of O3-N and O3-S with two radio sources, we do not find evidence of shocks from a radio jet that could be powering the EELRs.
   We interpret this as O3-S being externally irradiated by the AGN, akin to the famous Hanny’s Voorwerp object in the local Universe. In addition, more classical line ratio diagnostics (e.g., \oiii/\hbeta vs [N\,\textsc{ii}]/\halpha) put the DSFG itself in the AGN region of the diagrams, and therefore suggest it to be the most probable AGN host. 
   These results showcase the ability of JWST to unveil obscured AGN at high redshifts.
   }


   \titlerunning{Extended \oiii in \sysname}
   \authorrunning{Solimano et al.}

   \maketitle
%
%________________________________________________________________

\section{Introduction}\label{sec:intro}

In the current paradigm of galaxy formation, the densest structures form in the most massive halos at high redshifts ($z>2$), at the junctures of cosmic web filaments of galaxies and neutral gas.
These structures are known as protoclusters, as they eventually evolve into massive galaxy clusters at $z<1$ \citep[e.g.,][]{Baugh1998EpochOfGalaxyFormation}.
Protoclusters are sites where active star formation, supermassive black hole (SMBH) accretion, and dynamical interactions trigger powerful feedback processes at large scales \citep{Overzier2016ProtoclusterReview}. 
The central regions of protoclusters can harbor dozens of galaxies within $\lesssim \SI{100}{\kilo\parsec}$ \citep[e.g.,][]{Oteo2018DRC, Miller2018SPT2349, Hill2020SPT2349}, with several of them hosting active galactic nuclei (AGN) and/or starbursts, leading to dramatic effects on the surrounding gas in the form of outflows, shocks, tidal debris, and ionized nebulae.

Protoclusters undergoing their most rapid phase of growth are commonly (though not always) signaled by a luminous quasar \citep[QSO; e.g.,][]{Shen2007SdssQsoClustering, Hennawi2015QuasarGiantLyaNebula, Decarli2019FirstHighZQsoOverdensity}, a high-redshift radio galaxy \citep[HzRG; e.g.,][]{Venemans2007ProtoclustersHighzRadioGals, MileyAndDeBreuck2008HzrgsAndEnvironments, Wylezalek2013SpitzerRadioLoudAgn, Noirot2018StructuresCarlaSurvey}, and/or one or more submillimeter-bright dusty star-forming galaxies \citep[DSFGs; e.g.,][]{Riechers2014AlmaCiiAztec3, Casey2016UbiquityOfCoevalStarbursts, Hill2020SPT2349, Wang2021OverdensitiesSubmmSpt}. These sources are often embedded in giant \hi Lyman-$\alpha$ (\lya) nebulae, which in some cases reach scales of hundreds of kiloparsecs \citep[e.g.,][]{McCarthy1987ExtendedLyaEmissionRadioGal,Reuland2003GiantLyaNebulaeRadioGal, Borisova2016MUSELyaBlobsAroundQSOs, Swinbank2015MappingGiantLyaHaloRadioGal, Kikuta2019LyaViewHyperluminousQso, Guaita2022AzTec3Lya, Apostolovski2024ExtendedLyaSpt2349}.

The gas in such environments is known to be multiphase, and hence the extended emission is not restricted to \lya. Recent detections of extended CO, [C\,\textsc{i}] and \cii emission imply the existence of cold gas reservoirs tracing  widespread star formation and accretion \citep[e.g.,][]{Emonts2018SpiderwebCircumgalacticCi, Emonts2023CarbonCosmicStream, Umehata2021CiiObsOfLAB1}. Similarly, \heii-, \civ-, and \oiii-emitting ionized nebulae, are typically found to trace outflows and photoionization by AGN \citep[e.g.,][]{Overzier2013OpticalEmLinesLyaBlobB1, Cai2017ElaneInOverdensity}. \oiii nebulae are particularly common around HzRGs, where kinetic feedback also plays a role, as suggested by their alignment with the radio jets \citep[e.g.,][]{Nesvadba2017SinfoniRadioGalSurvey}.

The James Webb Space Telescope (JWST) is becoming an important tool to understand ionized nebulae within protoclusters, since it has opened access to the diagnostic-rich rest-frame optical spectrum at $z>3$. The Near-InfraRed Spectrograph's Integral Field Unit (NIRSpec IFU), in particular, has allowed the community to identify and characterize extended \oiii nebulae around quasars \citep[e.g.,][]{Wylezalek2022NirspecExtremelyRedQso, Perna2023GaNifsDualAgnEnvironment, Decarli2024NIrspecResdshiftSixQso}, HzRGs \citep[e. g.,][]{Saxena2024WidespreadFeedbackProtoBcgJwst, Roy2024JwstFeedbackRadioJets, Wang2024AgnIonizationConeRadioLoud}, and DSFGs \citep[e.g.,][]{PerezGonzalez2024NirspecJekyllHyde} in protoclusters or dense groups at high $z$.

In this paper, we present NIRSpec IFU observations of \sysname, a well-known $z=4.54$ galaxy group in the COSMOS field  \citep[e.g.,][]{Capak2008ExtremeStarburst, Smolcic2017SmgEnvironments} hosting a massive DSFG \citep[$M_{*}=\SI{8.7e10}{\msun}$,][]{Smolcic2015PhysicalPropertiesSmgsCosmos} and a luminous Lyman-break galaxy (LBG, $M_{UV}\approx-24.2$; \citealt{GomezGuijarro2018AlmaMinorMergersSmgs}) called CRISTAL-01a (hereafter C01) within the inner \SI{20}{\kilo\parsec}. This system resides in the center of a $z_\mathrm{phot}\approx4.5$ overdensity of LBGs at both small ($<\SI{2}{\arcmin}$) and large ($>\SI{2}{\arcmin}$) scales \citep{Smolcic2017SmgEnvironments, Jimenez-Andrade2023LyaNebulaeGalaxyPair}, and has been linked to the $z_\mathrm{spec}=4.57$ Taralay protocluster \citep[also known as PCI\,J1001+0220,][]{Lemaux2018VudsProtoCluster, Staab2024TaralayProtocluster}. Furthermore, \citet{Jimenez-Andrade2023LyaNebulaeGalaxyPair} observed \sysname using the Multi Unit Spectroscopic Explorer (MUSE) mounted on the Very Large Telescope (VLT), and found a  $L_{\mathrm{Ly}\alpha}\approx\SI{4e43}{\erg\per\second}$ \lya blob (LAB) and a handful of lower-mass \lya emitters distributed around the DSFG. Moreover, the authors confirm the results of \citet{Smolcic2017VlaCosmosEvolutionOfRadioAgn} using the COSMOS2020 catalog \citep{Weaver2022Cosmos2020}, and find an overdensity of $\delta_\mathrm{gal}=\num{6+-1}$ within a comoving volume of \SI{15}{\cubic\mega\parsec}. 

Three puzzling observations make \sysname an interesting case to study: first, the DSFG is detected at radio frequencies with $L_{1.4\,\mathrm{GHz}}=\SI{5.1+-1.2e24}{\watt\per\hertz}$ \citep{Carilli2008RadioLbgs, Capak2008ExtremeStarburst, Jimenez-Andrade2023LyaNebulaeGalaxyPair} that are possibly attributed to an AGN, yet have no X-ray counterpart. Secondly, the LAB is spatially and spectrally offset from the DSFG, but is coincident with the nearby LBG C01 \citep{Jimenez-Andrade2023LyaNebulaeGalaxyPair}.
Finally, \citet{Solimano2024CristalPlume} find a plume of \ciim line emission of \SI{15}{\kilo\parsec} in length toward \sysname using deep ALMA observations, indicating a dynamically complex system, although its physical origin remains unclear. The observations presented here reveal additional features that bring us closer to obtaining a full picture of the baryonic cycle around \sysname.

Throughout the paper, we assume a flat cosmology described by $H_0=\SI{70}{\kilo\meter\per\second\per\mega\parsec}$ $\Omega_{m,0}=0.3$, and $\Omega_{\Lambda, 0}=0.7$. At $z=4.54$, the physical scale is \SI{6.578}{\kilo\parsec\per\arcsec}.

\begin{figure*}[!htb]
    \centering
    \includegraphics[width=17cm]{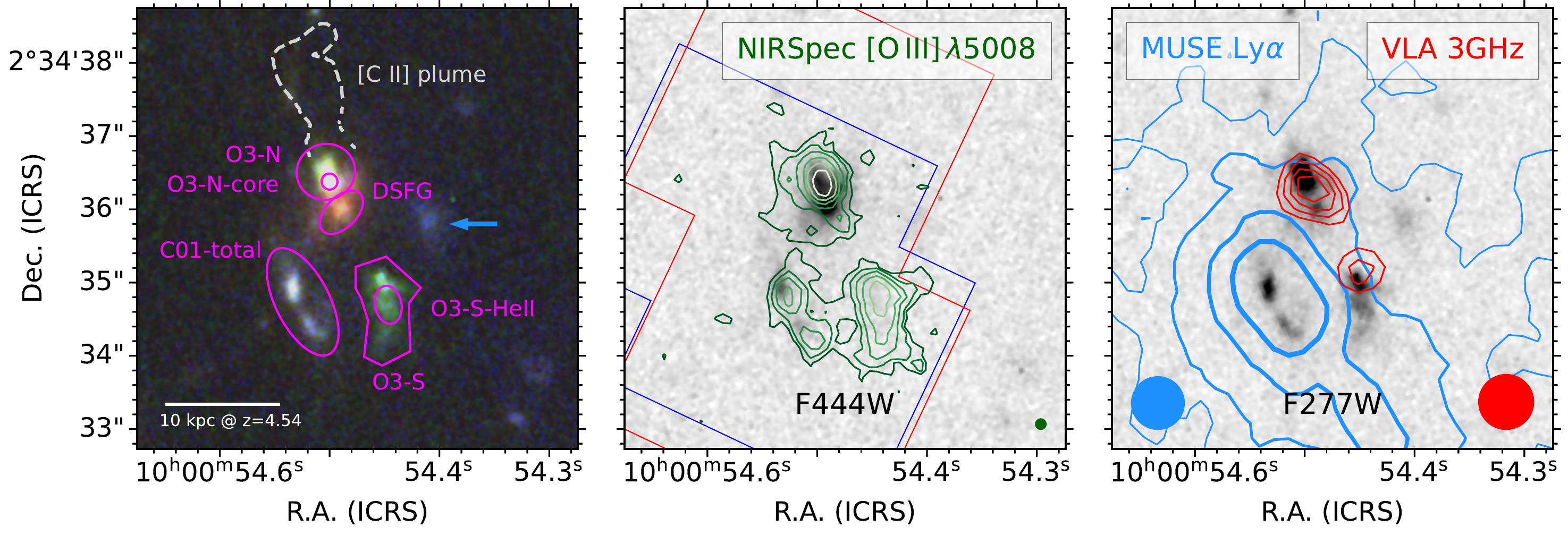}
    \caption{Multiwavelength view of the \sysname system at $z=4.54$.{\it Left panel:} Color composite image of the \sysname system as seen by JWST/NIRCam. The filters F356W, F277W, and F200W are mapped to the red, green, and blue channels, respectively. The F277W filter captures the \hbeta+\oiii emission at $z=4.54$. Magenta regions indicate the apertures used in this paper. The gray dashed contour delineates the \cii plume detected with ALMA \citep{Solimano2024CristalPlume}. The blue arrow points to a foreground galaxy at $z_{\mathrm{spec}}=1.41$ \citep{Capak2008ExtremeStarburst}. {\it Middle panel:} NIRCam F444W image with contours of the \oiii emission detected in the NIRSpec G235M observations. Contours start at $2\sigma=\SI{3.3e-17}{\erg\per\second\per\centi\meter\squared\per\arcsec\squared}$ and increase as integer powers of 2. The blue and red solid regions show the footprint of the observations obtained with G235M and G395H gratings, respectively. The green circle in the bottom right corner indicates the angular resolution element. \textit{Right panel:} NIRCam F277W image with $\pm3,\,4, 5$, and $6\sigma$ contours from Very Large Array (VLA) S-band continuum (red), and $\left\lbrace1,5,16,30\right\rbrace\times\SI{e-18}{\erg\per\second\per\centi\meter\squared\per\arcsec\squared}$ contours of \lya emission (blue) from VLT/MUSE \citep{Jimenez-Andrade2023LyaNebulaeGalaxyPair, Solimano2024CristalPlume}. Blue and red circles indicate the angular resolutions of the MUSE and VLA datasets, respectively.}
    \label{fig:nircam-rgb}
\end{figure*}

%________________________________________________________________

\section{Observations and data reduction}\label{sec:data}

\subsection{JWST/NIRCam data}\label{sec:data:nircam}
Multiband NIRCam imaging data of the \sysname system comprise a total of six broadband filters. Images using the F115W, F150W, F277W, and F444W filters were taken as part of the public Cosmos-Web survey \citep[GO-1727, PI: Kartaltepe \& Casey,][]{Casey2023CosmosWebOverview} using integration times of \SI{515}{\second} per filter at the position of \sysname, while the F200W and F356W bands were observed for \SI{1074}{\second} as part of GO-4265 (PI: Gonz\'alez-L\'opez).
At the redshift of our source, the F277W and F356W filters cover the [O\,\textsc{iii}]$+\mathrm{H}\beta$ and \halpha emission lines, respectively.

We reduced these data using the CRAB.Toolkit.JWST\footnote{\url{https://github.com/1054/Crab.Toolkit.JWST}} wrapper of the JWST pipeline (version 1.10.0, pmap=1075) with highly optimized parameters. In addition, we followed \citet{Bagley2023CeersNircamReduction} for 1/$f$ noise mitigation, applied background subtraction via the skymatch method of the standard pipeline, removed wisp artifacts using published templates \citep{Bagley2023CeersNircamReduction}, and finally aligned our images to the COSMOS2020 catalog \citep{Weaver2022Cosmos2020}. The combined images are drizzled to a common grid with a pixel size of \ang{;;0.02}.

\subsection{JWST/NIRSpec data}\label{sec:data:nirspec}
In this work, we used JWST/NIRSpec IFU data from programs GO-3045 (PI: Faisst) and GO-4265 (PI: González-Lopez) that target the \sysname system with the G235M ($\SI{1.7}{\micro\meter}<\lambda<\SI{3.2}{\micro\meter}$, $R\sim1000$) and G395H ($\SI{2.9}{\micro\meter}<\lambda<\SI{5.3}{\micro\meter}$, $R\sim2700$) gratings, respectively. 

The G235M dataset was taken using two \SI{1080}{\second} dithered exposures with overlap at the location of C01. The G395H dataset was set up as a two-tile mosaic covering both C01 and the \cii plume reported by \citet{Solimano2024CristalPlume}. Each tile was observed for \num{5974} seconds.

The data were reduced with the standard JWST pipeline (version 1.12.5, pmap=1234) plus some additional tweaks. Briefly, we followed the scripts provided by \citet{Rigby2023TemplatesOverview}\footnote{Available at \href{https://zenodo.org/doi/10.5281/zenodo.10737011}{10.5281/zenodo.10737011}} but implemented improved snowball removal in Stage 1, and additional bad pixel flagging after Stage 1. Also, we switched on the outlier-rejection step in Stage 3, and turned off the master background subtraction. Instead, background subtraction was performed as a post-processing step, together with stripe mitigation and astrometric alignment to NIRCam. A more detailed description of the reduction is presented elsewhere (Fujimoto et al., in prep.).

%________________________________________________________________
\section{Results and analysis}\label{sec:results}

The NIRCam images reveal significant emission from several sources that were faint in previous Hubble Space Telescope (HST) imaging \citep{GomezGuijarro2018AlmaMinorMergersSmgs, Solimano2024CristalPlume}.
For example, the DSFG starlight is now clearly detected in the long-wavelength filters.
Interestingly, two other sources dominate the emission in the F277W and F356W filters (appearing green in Fig. \ref{fig:nircam-rgb}), indicating the possibility of high-equivalent-width \oiii, \hbeta, and \halpha emission lines.
The first of these sources is just \ang{;;0.5} north of the DSFG, at the same location as an HST source \citep[\sysname-North in][O3-N hereafter]{GomezGuijarro2018AlmaMinorMergersSmgs}.
The other is located south of the DSFG (hence O3-S), and has a projected extent of $\ang{;;1.3}=\SI{8.55}{\kilo\parsec}$, and extremely faint HST magnitudes ($m_\mathrm{F125W}\approx26$ AB). 

JWST/NIRSpec observations confirm the presence of strong \oiii emission at the locations of O3-N (EW$_0=\SI{1780+-80}{\angstrom}$) and O3-S (EW$_0=\SI{5100+-1000}{\angstrom}$)\footnote{The equivalent width values presented here consider only the \SI{5008}{\angstrom} line of the \oiii doublet.} and, more importantly, at the same redshift as \sysname, therefore confirming their physical association (see the middle panel of Fig.~\ref{fig:nircam-rgb}). The nebulae also seem to be co-spatial with the \SI{3}{\giga\hertz} radio detections, but they are offset from the \lya peak surface brightness (SB; see the right panel of Fig.~\ref{fig:nircam-rgb}).

In the following subsections, we use apertures to extract and explore the spectroscopic properties of the two \oiii nebulae.  The labeled apertures in Fig.~\ref{fig:nircam-rgb} were manually defined based on the RGB NIRCam image and the \oiii map. For the DSFG we used an aperture significantly smaller than the full extent of the source to avoid contamination from O3-N. We also defined two sub-apertures within the \oiii nebulae that either enclose the peak of \oiii emission (O3-N-core) or maximize the signal-to-noise ratio (S/N) of the \heii line (O3-S-HeII).

\subsection{Morphology and kinematics}\label{sec:results:morphokin}

\begin{figure}[!hbt]
    \centering
    \resizebox{\hsize}{!}{\includegraphics{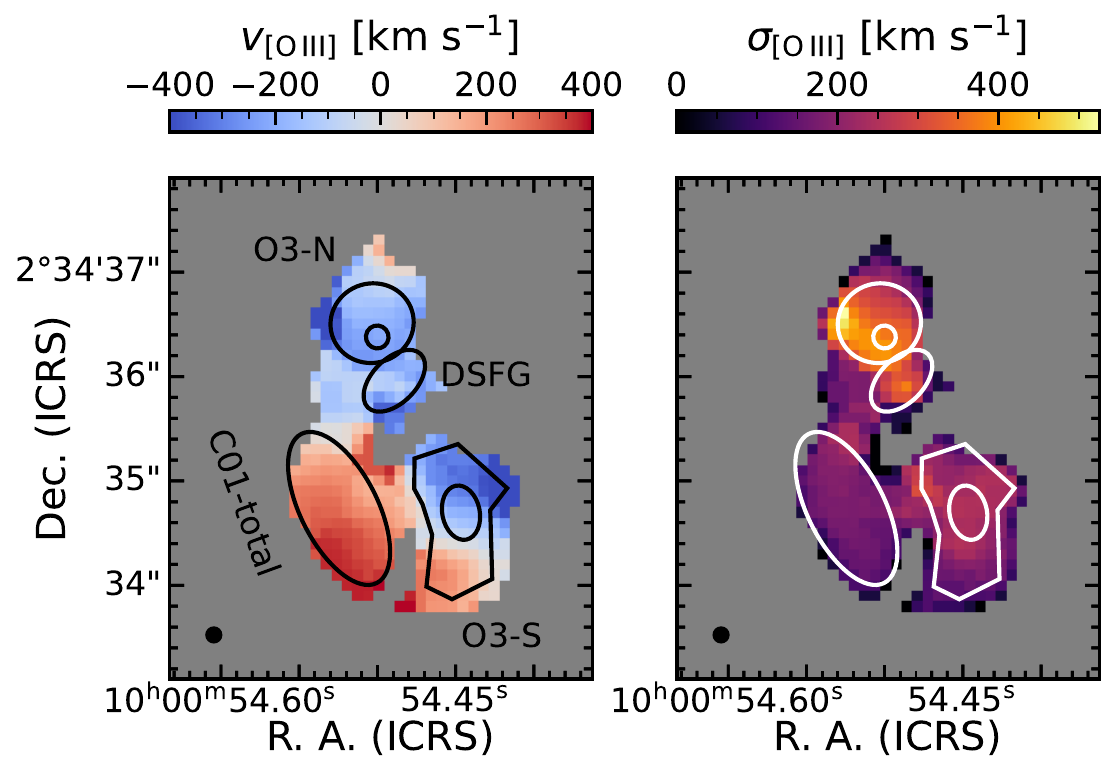}}
    \caption{Resolved \oiii kinematics of \sysname. Velocity field (left) and velocity dispersion (right) maps. The reference velocity is defined at $z=4.5471$.}
    \label{fig:o3-moments}
\end{figure}

The middle panel in Fig. \ref{fig:nircam-rgb} shows the distribution of \singlet{O}{iii}{5008} SB around \sysname.
As expected, significant emission was detected in O3-N and O3-S, but also on C01 and the DSFG.
Moreover, the global \oiii emission seems to be spatially extended and low SB emission connects nearly all of the objects in the scene.

Figure \ref{fig:o3-moments} features the velocity field and velocity dispersion maps of the \oiii emission line in the system.
These maps were created following \citet{Solimano2024CristalPlume}, with a spatial and spectral Gaussian convolution kernel applied to the continuum-subtracted cube.
The spatial kernel has $\sigma=1$ spaxel, whereas the spectral kernel has $\sigma=\sigma_{LSF}$ at the wavelength of the line. 
The moments were created by masking out all the voxels with S/N$<3$ in the convolved cube.
In the velocity field map we see the DSFG and O3-N share similar velocities, with an offset of $\approx \SI{500}{\kilo\meter\per\second}$ with respect to C01. 
In turn, O3-S shows a large velocity gradient north-to-south, with a velocity span of almost $\sim \SI{800}{\kilo\meter\per\second}$ from end to end.
If we were to interpret this gradient as a signature of virialized rotation, a rough calculation would yield a dynamical mass on the order of $Rv^2/G=(\SI{4.3}{\kilo\parsec})(\SI{400}{\kilo\meter\per\second})^2 / G \approx \SI{1.6e11}{\msun}$.
This value is comparable to the dynamical mass of the DSFG \citep{Fraternali2021FastRotators}, but since O3-S lacks significant stellar or dust emission, we deem unlikely that O3-S is a massive rotator.
Instead, O3-S could be tidal debris from an ongoing interaction between the members of the system. In particular, the presence of a low-SB bridge between O3-S and C01, together with matching line-of-sight velocities in the southern end of both sources, already hints at a tidal origin. Further discussion of this scenario is presented in Sect.~\ref{sec:discussion}.

Additionally, the velocity dispersion map of the \oiii emission shows a fairly uniform structure at \SI{200}{\kilo\meter\per\second} in most of the system except for O3-N.
The velocity dispersion in O3-N reaches \SI{500}{\kilo\meter\per\second}, indicating a higher dynamical mass, increased turbulence, or additional kinematic components. 

\subsection{Broad velocity component in O3-N}\label{sec:results:broad}

Inspection of the \oiii and \halpha line profiles in the O3-N aperture reveal the presence of broad velocity wings.
To characterize this additional kinematic component, we fitted single and double Gaussians plus a constant continuum level as detailed in Appendix~\ref{sec:ap:gaussfit}.

 \begin{figure*}[!htb]
     \centering
     \includegraphics[width=17cm]{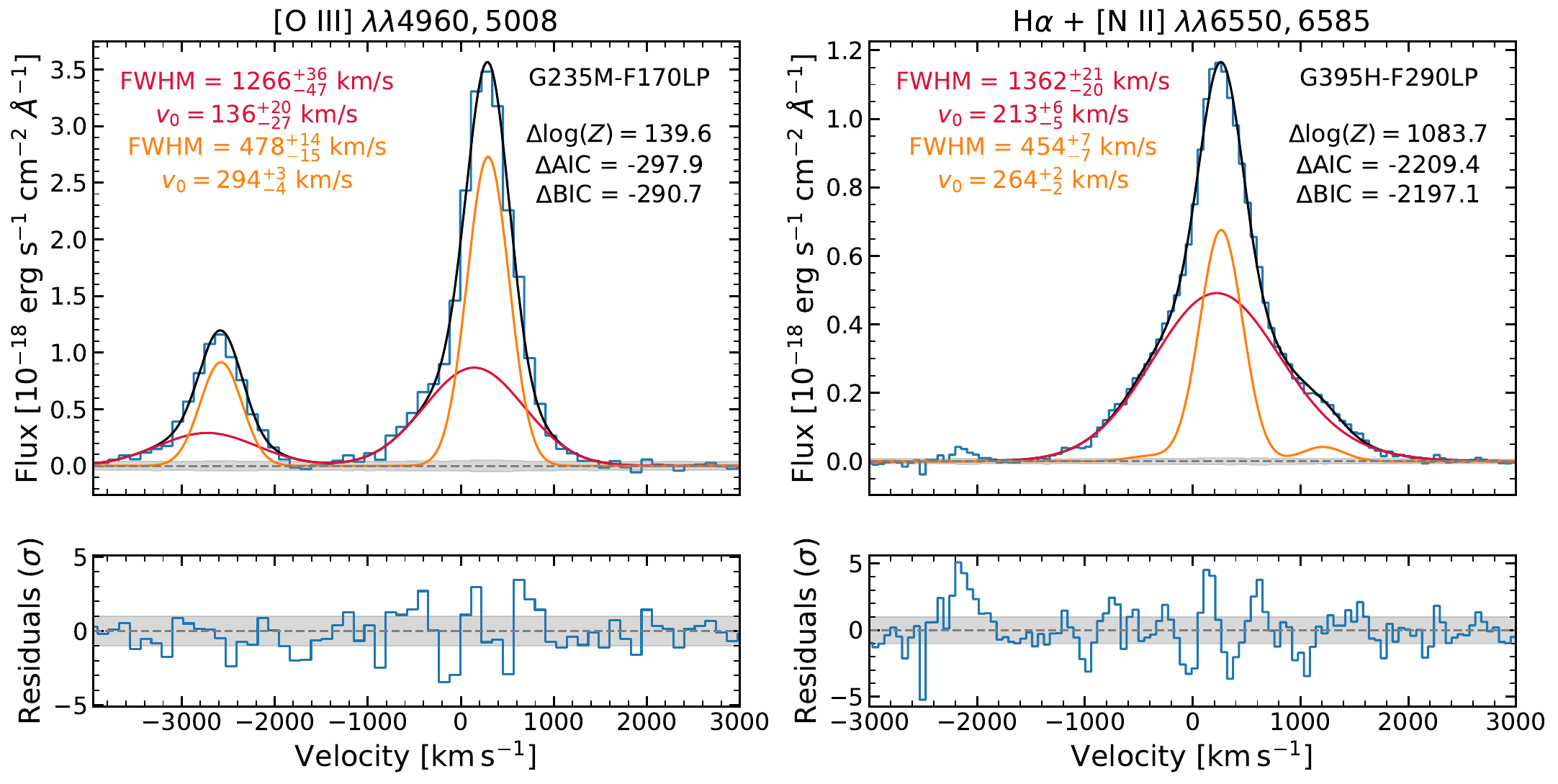}
     \caption{Evidence of a broad velocity component in the strongest emission lines from O3-N, namely \doublet{O}{iii}{4960}{5008} (left) and \halpha+\doublet{N}{ii}{6550}{6585} (right). In both lines, a double Gaussian profile is preferred over a single Gaussian fit (not shown), based on the former having a larger $\log\left (Z\right)$, and  lower AIC and BIC scores (see upper right legends). The bottom panels show the residuals of the subtraction of the best-fit model from the data, in units of $\sigma$.}
     \label{fig:o3n-broad}
 \end{figure*}

The results of our fits are shown in Figure~\ref{fig:o3n-broad}. The double Gaussian model is preferred over the single one based on its higher Bayesian evidence score, and lower Akaike Information Criterion  \citep[AIC; e.g.,][] {Cavanaugh1997CorrectedAkaikeCriterion} and Bayesian Information Criterion \citep[BIC; e.g.,][]{Schwarz1978BayesianInfoCriterion} scores.

 \begin{figure*}[!htb]
     \centering
     \includegraphics[width=17cm]{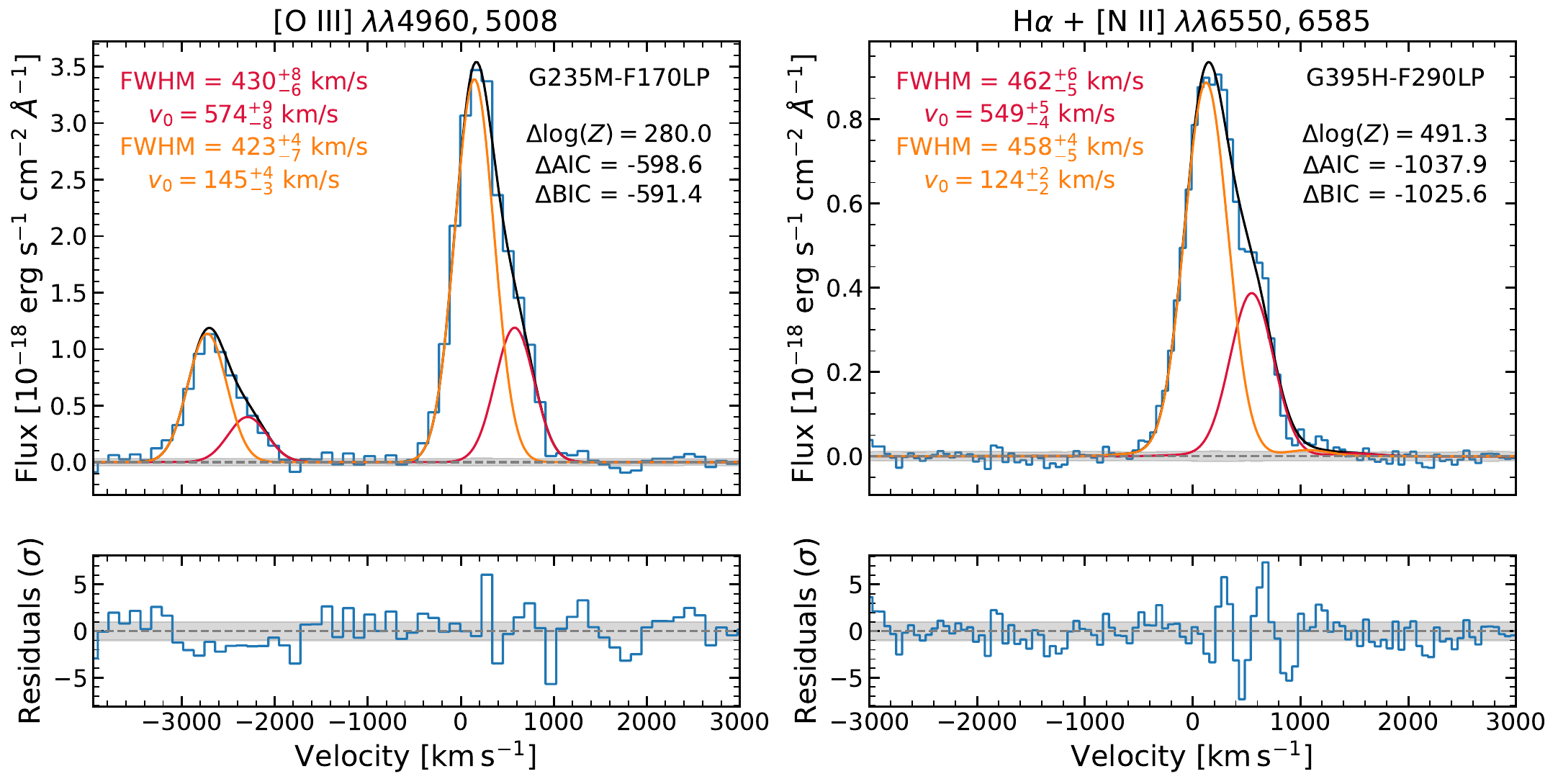}
     \caption{Same as Fig.~\ref{fig:o3n-broad} but for O3-S. Here, both components have similar line width, in contrast with the expectations for an outflow. }
     \label{fig:o3s-broad}
 \end{figure*}

The \oiii broad component of O3-N displays a full width at half maximum (FWHM) of $1266_{-47}^{+36}\,\si{\kilo\meter\per\second}$, and is blueshifted by \SI{158+-24}{\kilo\meter\per\second} from the central velocity of the narrow component. Such a profile of the \oiii line (broad and blueshifted) typically points to the existence of strong ionized outflows projected onto the line of sight. \halpha shows an even broader but less blueshifted profile.
%The rationale behind this comes from considering simple bipolar outflow geometries, where the receding (redshifted) side is partially behind the disk of the galaxy, thus suffering from increased dust attenuation and \oiii self-absorption \comment{cite!!}.

Notably, since O3-N sits at the base of the \cii plume (see left panel of Fig.~\ref{fig:nircam-rgb}) and has a broad \oiii component at the same velocity $(v_0\approx\SI{150}{\kilo\meter\per\second})$ as the corresponding \ciim line, the outflow scenario proposed by \citet{Solimano2024CristalPlume} emerges as a natural explanation. A detailed assessment of this possibility will be presented in a forthcoming paper.

The spectrum of O3-S (Fig.~\ref{fig:o3s-broad}) also shows a secondary velocity component, but this likely arises from the large velocity gradient (cf. Fig.~\ref{fig:o3-moments}) contained within the aperture, or by the superposition of two nebulae separated by roughly \SI{400}{\kilo\meter\per\second}. An outflow origin for O3-S seems less plausible because the two components have the same velocity width.

\subsection{Line ratio diagnostics and high-ionization species}\label{sec:results:bpt}

\begin{figure*}[!hbt]
    \centering
    \includegraphics[width=17cm]{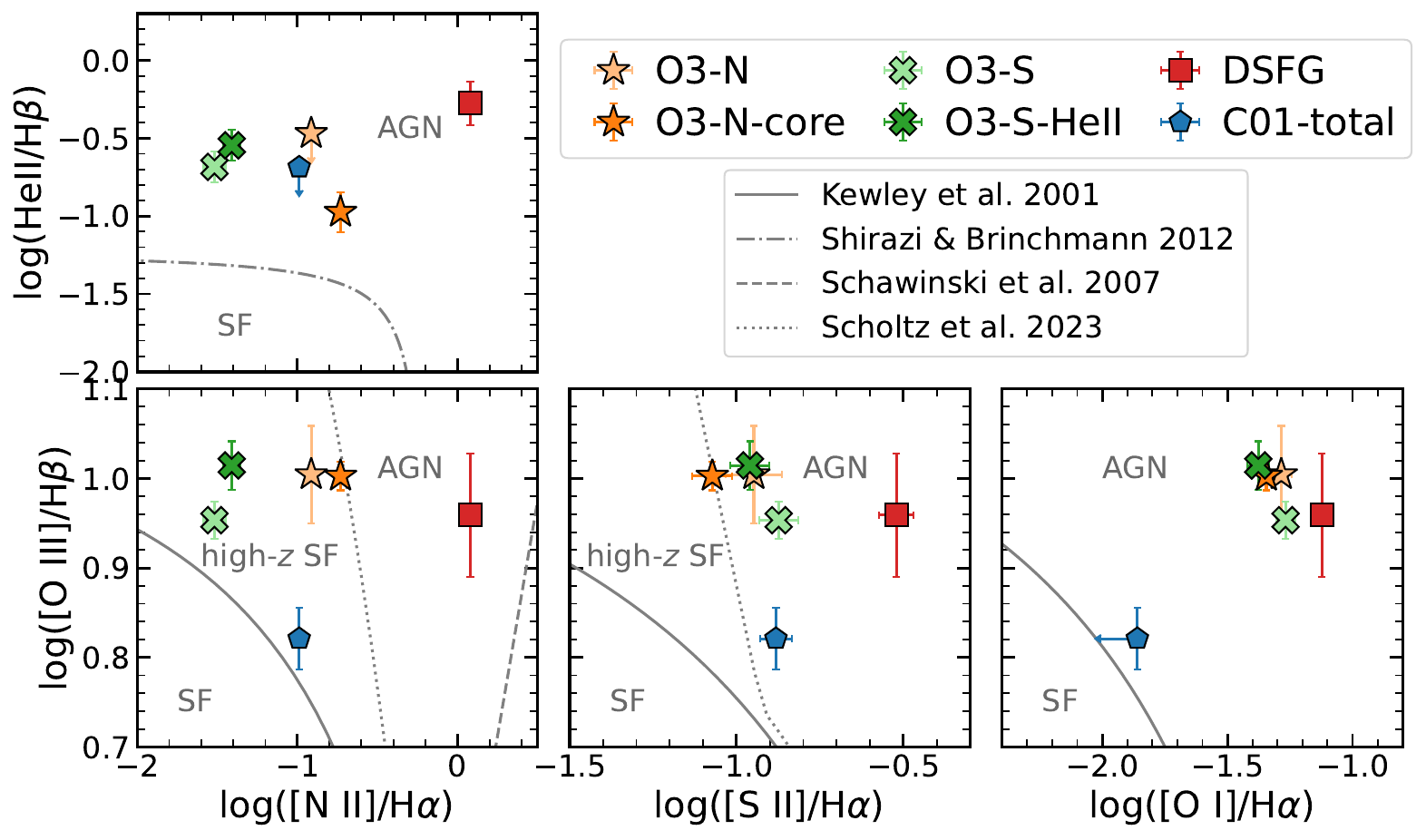}
    \caption{Diagnostic line ratio diagrams for the relevant regions of the \sysname system. \textit{Bottom left:} Standard \citetalias{Baldwin1981BPTpaper} R3 vs N2 diagram. The solid curved lines represent the maximum starburst line of \citet{Kewley2001ModelingStarbursts}, while the straight dashed line on the lower right-hand side divides AGN from LINERs according to \citet{Schawinski2007AgnFeedbackSdssLiners}. The dotted line is the separation from SF and AGN adapted for high-$z$ galaxies by \citet{Scholtz2023JadesLargeAgnPop}. \textit{Bottom center:} R3 vs S2 \citetalias{VeilleuxAndOsterbrock1987SpectralClassGalaxies} diagram. Again, the solid line is from \citet{Kewley2001ModelingStarbursts} and the dotted line is the high-$z$ SF/AGN separation from \citet{Scholtz2023JadesLargeAgnPop} \textit{Bottom right:} R3 vs O1 \citetalias{VeilleuxAndOsterbrock1987SpectralClassGalaxies} diagram. The solid line separates SF from AGN according to \citet{Kewley2001ModelingStarbursts}. \textit{Top left:} He2 vs N2 diagram. The division line between AGN and SF is taken from \citet{ShiraziAndBrinchmann2012SdssHeIIemission}.}
    \label{fig:regions-bpt}
\end{figure*}

We measured all line fluxes and errors using \verb|pPXF| \citep{Cappellari2017FullSpectrumFitting, Cappellari2023PpxfUpdate} as detailed in Appendix~\ref{sec:ap:ppxf}. From these, we computed the five line ratios presented in Fig.~\ref{fig:regions-bpt}. 
The bottom panels of Fig.~\ref{fig:regions-bpt} show three diagrams displaying the R3=\singlet{O}{iii}{5008}/\hbeta ratio against three different line ratios, namely N2=\singlet{N}{ii}{6583}/\halpha \citep*[][the ``BPT'' diagram]{Baldwin1981BPTpaper}, S2=\doublet{S}{ii}{6716}{6731}/\halpha, and O1=\singlet{O}{i}{6302}/\halpha \citep[also known as the][or VO87 diagrams]{VeilleuxAndOsterbrock1987SpectralClassGalaxies}.
In each of them we plotted the theoretical boundary between star-formation (SF) and AGN photoionization models from \citet{Kewley2001ModelingStarbursts} and \citet{Kewley2006AgnClassification}.
In the \citetalias{Baldwin1981BPTpaper} diagram we also plotted the boundary between AGN and low-ionization nuclear emission regions \citep[LINERs;][]{Schawinski2007AgnFeedbackSdssLiners}.
Recent JWST observations have found that these classical boundary lines are not reliable at the low metallicities and high-ionization conditions typically found at high redshifts, which make SFGs and AGN overlap in these diagrams \citep[e.g.,][]{Harikane2023NirspecFirstAgnCensus, Maiolino2023JadesPopulationBlackHoles, Scholtz2023JadesLargeAgnPop}.
Hence, we also plotted the more conservative boundary line proposed by \citet{Scholtz2023JadesLargeAgnPop} to account for this issue. In addition, we show in Fig.~\ref{fig:regions-bpt} the He2-N2 diagram \citep{ShiraziAndBrinchmann2012SdssHeIIemission}, featuring He2=\permitted{He}{ii}{4686}/\hbeta vs N2. This diagnostic diagram has been cited as a more robust way to select AGN than the classical \citetalias{Baldwin1981BPTpaper} and \citetalias{VeilleuxAndOsterbrock1987SpectralClassGalaxies} diagrams \citep[e.g.,][]{Scholtz2023JadesLargeAgnPop}, although the \permitted{He}{ii}{4686} is often too faint to be detected.

Despite these caveats, we see in Fig.~\ref{fig:regions-bpt} that only C01 appears to be consistent with pure SF, while the rest can be explained at least partially by AGN excitation. On the other hand, the DSFG is the only source showing AGN-like ratios in all diagrams.

The possible presence of an AGN has already been proposed by \citet{Jimenez-Andrade2023LyaNebulaeGalaxyPair} based on the radio detection, and the \permitted{He}{ii}{1640}/\lya and \permitted{C}{iv}{1551}/\lya ratios. Additional support to this idea comes from the detection of the \singlet{Ne}{v}{3427} line toward O3-N-core (see Fig.~\ref{fig:nev-O3-N-core}), since [Ne\,{\sc v}] requires photons with $E>\SI{97.11}{\electronvolt}$. Such high energies are most easily attainable with AGN activity, either in the form of photoionization or fast shocks \citep[e.g.,][]{Gilli2010XrayToNevRatio, Mignoli2013ObscuredAgnZcosmos, Leung2021KcwMrk273, Cleri2023Ne53RatioEig}.

In O3-S and O3-S-HeII, at the other side of the DSFG, we did not detect \singlet{Ne}{v}{3427}, but only \permitted{He}{ii}{4686} (see Fig.~\ref{fig:nev-O3-S-HeII}). Due to the lower ionization energy of helium ($E>\SI{54.42}{\electronvolt}$), this line is not as clean an indicator of AGN as the [Ne\,\textsc{v}] line, and can indeed be excited by X-ray binaries \citep[e.g.,][]{Schaerer2019XrayBinariesOriginHeII}, Wolf-Rayet stars \citep[e.g.,][]{ShiraziAndBrinchmann2012SdssHeIIemission}, and shocks \citep{Izotov2012NevEmissionBlueCompactDwarfs}, among others. Nevertheless, its location on the He2-N2 diagram is well above the SF boundary line.

\begin{figure*}[!hbt]
  \centering
  \includegraphics[width=17cm]{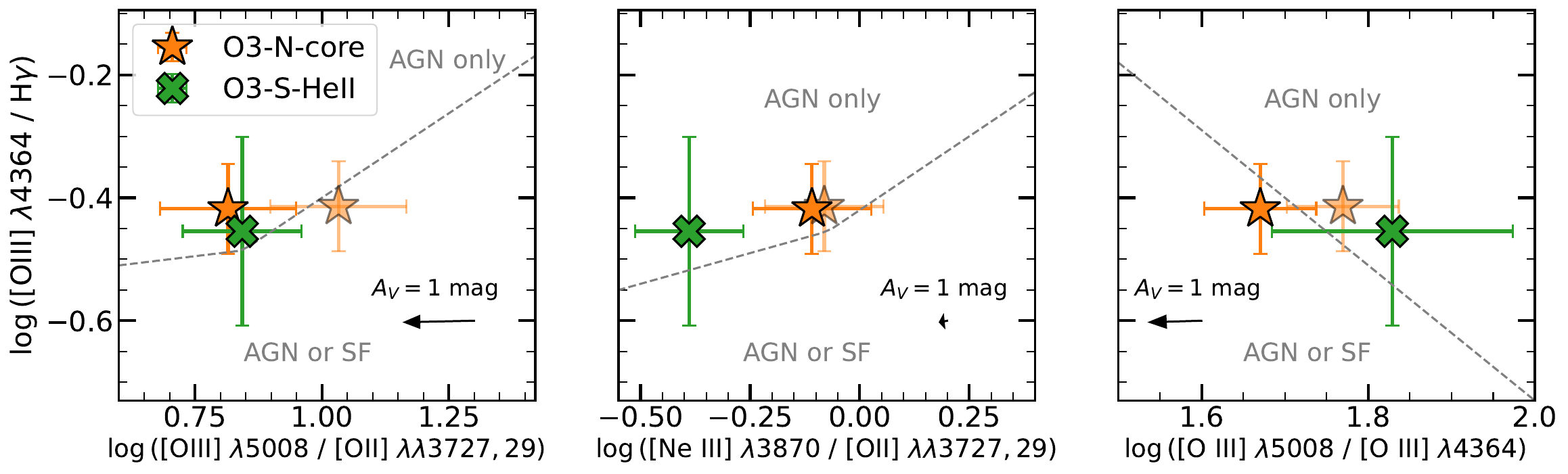}	
  \caption{Diagnostic line-ratio diagrams based on \singlet{O}{iii}{4364} emission. In all three panels the dashed line is the empirical demarcation between pure AGN and AGN/SF mixtures proposed by \citet{Mazzolari24NewAgnDiagnostic}. The faded markers with error bars are determined from our pPXF fits to the spectra without reddening correction. The bold markers show the values after reddening correction, with $A_V=\num{1.58+-0.16}$\,mag for O3-N-core and $A_V=0$ for O3-S-HeII. The black arrows denote the dereddening vector for an attenuation of $A_V=1$ mag with the \citet{Calzetti2000Dust} attenuation law.}\label{fig:mazzolari}
\end{figure*}

Finally, we also detected the temperature-sensitive \singlet{O}{iii}{4364} auroral line toward O3-N-core and O3-S-HeII (see Appendix~\ref{sec:ap:faint}).
A high ratio between \singlet{O}{iii}{4364} and the H$\gamma$ line was recently proposed as an alternative indicator of AGN photoionization \citep[e.g.,][]{Ubler2024OffsetAgnGaNifs}.
In Fig.~\ref{fig:mazzolari}, we plot the position of O3-N-core and O3-S-HeII in the three diagnostic diagrams developed by \citet{Mazzolari24NewAgnDiagnostic} based on \singlet{O}{iii}{4364} emission. We applied a reddening correction of $A_{V}=\num{1.58+-0.16}$\,mag to O3-N-core based on its Balmer decrement (see Sect.~\ref{sec:agnlum}), while O3-S-HeII is consistent with zero dust.
Our points nominally fall in the AGN-only region in two out of three diagrams, but they cross the boundary lines within the uncertainties. Therefore, we are unable to make any firm conclusions regarding the AGN nature of O3-N-core and O3-S-HeII based on the \citet{Mazzolari24NewAgnDiagnostic} diagrams.

\section{Oxygen abundance}\label{sec:metallicity}
We measured gas-phase oxygen abundances in the spectra of the different apertures using the indirect indicator proposed by \citet{Dopita2016MetallicityIndicator}. This indicator is calibrated as 
\begin{equation}
    12+\log(\mathrm{O}/\mathrm{H}) = 8.77 + \log\left(\mathrm{N II}/\mathrm{SII}\right) + 0.264\log\left(\mathrm{NII}/\mathrm{H}\alpha\right).
\end{equation}
We chose this indicator because it uses lines from a single grating/filter combination (G395H), thus avoiding possible systematic effects from the combination of the two datasets, and also because it is fairly robust to dust attenuation effects (which are significant at least in the case of the DSFG). The main caveat is that it relies on the assumption of a specific relation between N/O and O/H abundances.

\begin{figure}
     \centering
     \resizebox{\hsize}{!}{\includegraphics{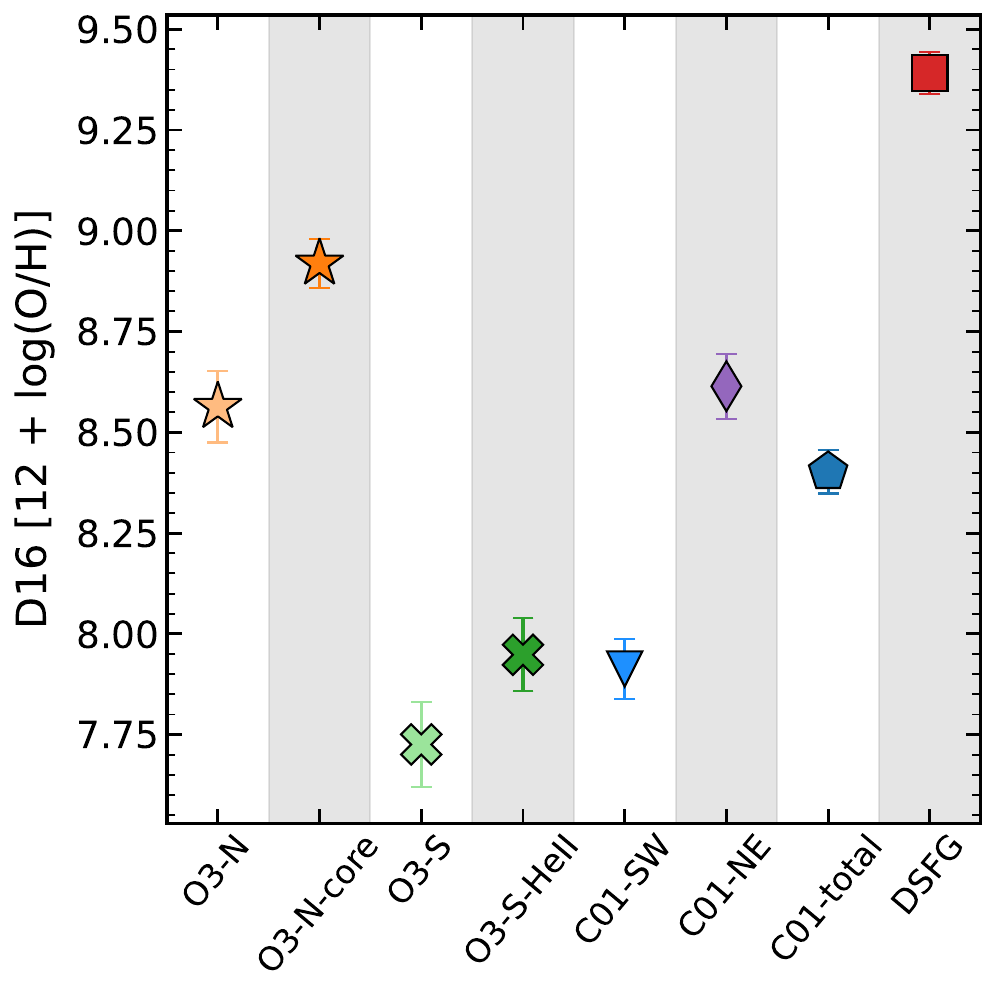}}
     \caption{Oxygen abundances of the sources in the \sysname system according to the \citet{Dopita2016MetallicityIndicator} calibration. Markers are the same as in Fig.~\ref{fig:regions-bpt}, except we also show C01-NE and C01-SW.}
     \label{fig:metallicities}
 \end{figure}

Figure \ref{fig:metallicities} shows the values obtained for all the apertures considered in this paper, including dedicated apertures for the two clumps C01-SW and C01-NE. As expected, the DSFG shows the highest (even supersolar) oxygen abundance. The rest of the apertures are distributed throughout the abundance scale, with O3-S showing the lowest abundance. We also report a large difference  ($\sim0.8$ dex) between C01-SW and C01-NE, with the latter dominating the integrated value (C01-total).

\section{Discussion}\label{sec:discussion}
We found in the previous section that the two strongest \oiii nebulae in the system are likely related to AGN activity. In the following subsections, we explore the possibility of radiative shocks, provide an estimate of AGN luminosity and obscuring column density given the current constraints, and then discuss the physical scenario for the origin of the \oiii nebulae.

\subsection{No evidence of shocks}\label{sec:shocks}

The detection of radio emission at the positions of O3-N and O3-S suggests the presence of shocks produced by a radio jet.
This might explain, for example, the enhanced velocity dispersion observed even in the ``narrow'' component of the emission lines ($\sigma \gtrsim \SI{100}{\kilo\meter\per\second}$).
However, the observed line ratios (see Fig.~\ref{fig:regions-bpt}) do not resemble those expected in fast radiative shocks, in contrast to those observed in the ``Ulema'' galaxy, a low-mass radio-detected AGN at $z=4.6$ \citep{DEugenio2024NirspecWideShockAgn}.
In particular, shock models assuming solar abundances predict N2, S2 and O1 ratios  that are typically above $-0.5$ dex, $-0.5$ dex, and $-1.2$ dex, respectively \citep[e.g.,][]{Allen2008Mappings3ShockLibrary, AlarieMorisset2019Mappings5ShockDb}.
At lower metallicities, these ratios can approach the values we observe in our data, but the R3 ratio decreases as well.

These results indicate that shocks play a negligible role in the ionization of the O3-N and O3-S nebulae. To further test this idea, we explored the relationship between the shock-sensitive N2 ratio and the line width. 
In shocked gas, a positive correlation has been found between N2 and velocity dispersion \citep[e.g.,][]{Rich2011GalaxyWideShocksLirgs, Ho2014SamiShocksAndOutflows, Rich2015MergersDriveShocksGoals}, indicating a coupling between the gas ionization and kinematics, which is not predicted in pure photoionization models \citep{Kewley2019EmissionLinesReview}.

We thus took advantage of JWST NIRSpec IFU's spatial resolution to measure the resolved N2 ratio within O3-N and O3-S.
We extracted spectra from the 30 and 39 spaxels contained in O3-N and O3-S apertures (see Fig~\ref{fig:nircam-rgb}), respectively.
We then fitted each spectrum with single and double Gaussians using the method described in Appendix~\ref{sec:ap:gaussfit}. 
Finally, we removed all fits where the posterior error on log(\singlet{N}{ii}{6585}/\halpha) is larger than \num{0.3} dex. 

\begin{figure*}[!htb]
  \centering
  \includegraphics[width=17cm]{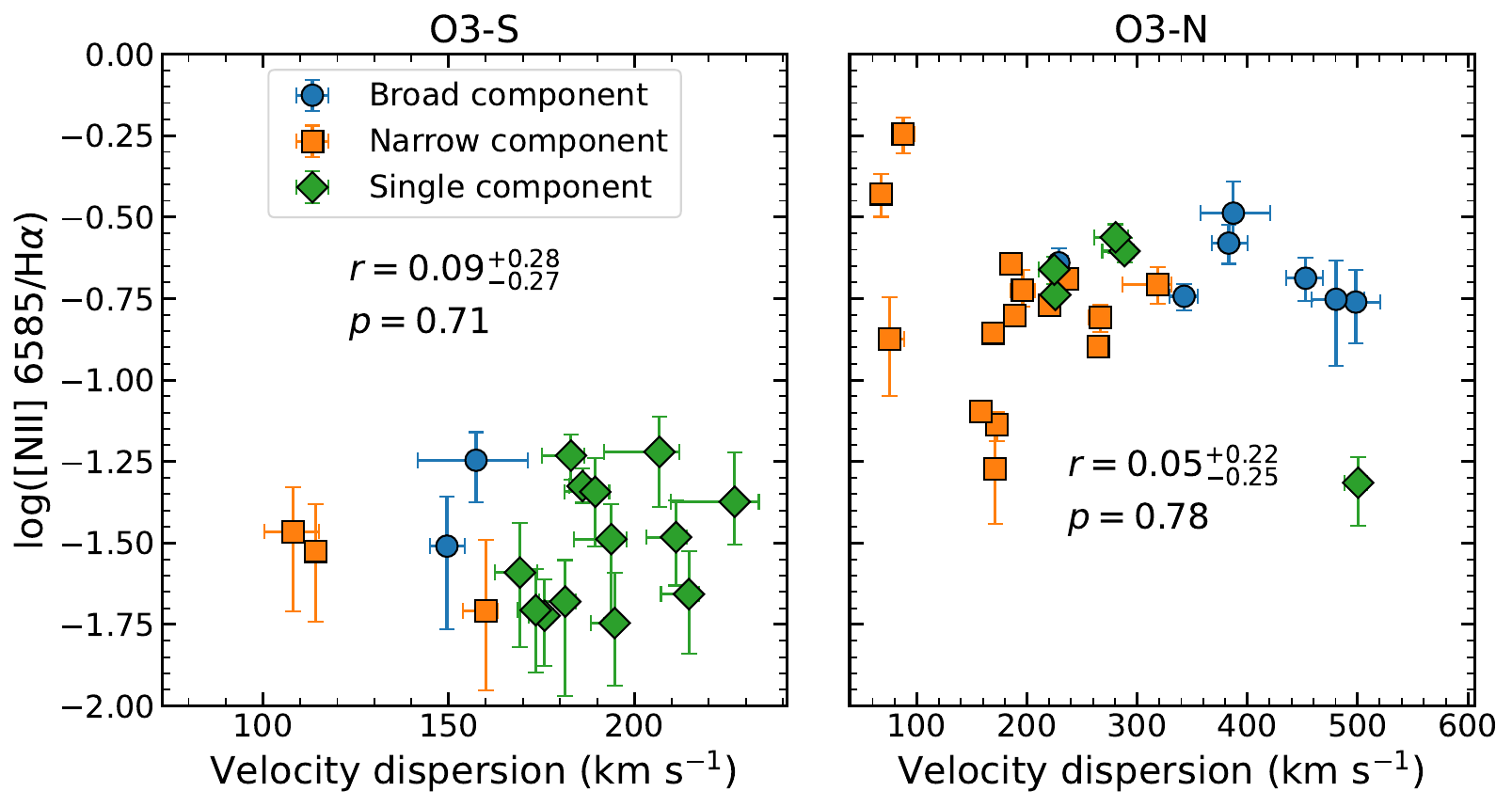}
  \caption{Resolved N2 ratio as a function of line velocity dispersion in O3-S (left) and O3-N (right). The inset text displays the Spearman's rank correlation coefficient $r$ and its corresponding $p-$value. No significant correlation is present in either panel.}\label{fig:resolved_nii_to_ha}
\end{figure*}

Obtained values of N2 and velocity dispersion ($\sigma$) where the best fit is a single Gaussian component are shown with green markers in Fig.~\ref{fig:resolved_nii_to_ha}, while the results where a double component is preferred are shown in orange and blue markers for the narrow and broad components, respectively.

We computed the Spearman's rank correlation coefficient independently for each set of spaxels, and estimated the uncertainties from 600 bootstrap samples.
We find no significant correlation between the line ratio and the velocity dispersion in either O3-N ($r={0.05}_{-0.25}^{+0.22}$, $p=0.78$) or O3-S ($r=0.09_{-0.27}^{+0.28}$, $p=0.71$). Instead, the N2 ratio remains relatively uniform across the dispersion axis, with a median value of $-0.74$ dex in O3-N and $-1.5$ dex in O3-S. The different median ratio between the two nebulae could be explained by different metallicities (see Sec.~\ref{sec:metallicity}).

\subsection{Constraints on AGN luminosity and obscuration}\label{sec:agnlum}
As pointed out by several authors \citep[e.g.,][]{Capak2008ExtremeStarburst, Smolcic2015PhysicalPropertiesSmgsCosmos, Jimenez-Andrade2023LyaNebulaeGalaxyPair}, \sysname was undetected in \textit{Chandra}'s \SI{80}{\kilo\second} observations in the \SIrange{0.5}{2}{\kilo\electronvolt} band \citep{Elvis2009ChandraCosmosOverview}, leading to a flux upper limit of \SI{3e-16}{\erg\per\second\per\centi\meter\squared} \citep{Capak2008ExtremeStarburst}. Assuming a Galactic foreground column density of $N_{H}=\SI{2.6e20}{\per\centi\meter\squared}$ and a power-law source with a photon index of $\Gamma=1.4$, this upper limit translates to a rest-frame hard X-ray luminosity of $L_{2-10\mathrm{keV}}< \SI{6.8e43}{\erg\per\second}$ \citep{Solimano2024CristalPlume}.
%Then, using a bolometric correction of 10 \citep{VasudevanAndFabian2007BolCorrAGN}, we infer a luminosity no larger than $L_\mathrm{bol}<\SI{6.8e44}{\erg\per\second} $.

We then tried to determine whether such an AGN can explain the observed narrow-line luminosities.
To address this question, we used the scaling relations of \citet{Berney2015BassXrayOpticalLines} based on a sample of nearby, hard X-ray-selected AGN for \oiii and other optical lines as a function of $L_{2-10\mathrm{keV}}$.
In O3-N, we measured an \singlet{O}{iii}{5008} flux of \SI{2.51+-0.04e16}{\erg\per\second\per\centi\meter\squared}, and a Balmer decrement of $\mathrm{H}\alpha/\mathrm{H}\beta=\num{4.9+-0.6}$.
Assuming case B recombination with an intrinsic ratio of $\mathrm{H}\alpha/\mathrm{H}\beta=\num{2.86}$ and a \citet{Calzetti2000Dust} attenuation law, we derived $A_V=\num{1.9+-0.4}$ mag.
We thus inferred a reddening-corrected \singlet{O}{iii}{5008} luminosity of \SI{2.7+-0.5e44}{\erg\per\second}.

The \citet{Berney2015BassXrayOpticalLines} relation predicts that at an intrinsic $2-10\,\si{\kilo\electronvolt}$ luminosity of \SI{6.8e43}{\erg\per\second}, the \oiii luminosity reaches $\log(L_{[\mathrm{O}\,\mathrm{III}]}/[\si{\erg\per\second}])=42\pm0.6$. Despite the large scatter of the \citeauthor{Berney2015BassXrayOpticalLines} relation, our measured \oiii luminosity exceeds the prediction as a $4\sigma$ outlier. This means that either (1) the X-ray source is heavily obscured and thus the intrinsic X-ray luminosity is much larger, (2) the \oiii emission is not only excited by the AGN but is rather mainly excited by SF, (3), the \citet{Berney2015BassXrayOpticalLines} relation is not applicable at this redshift, or indeed (4) a combination of all of the above.

At face value, the measured \oiii luminosity would correspond to an intrinsic X-ray luminosity of $L_{2-10\,\si{\kilo\electronvolt}}=\SI{e46}{\erg\per\second}$, after extrapolating the range of the \citet{Berney2015BassXrayOpticalLines} relation, and assuming no contribution from SF.
We used the simple absorbed power-law model with $\Gamma=1.4$ within the Chandra PIMMS tool\footnote{\url{https://asc.harvard.edu/toolkit/pimms.jsp}} (version 4.12d) to infer a column density of $N_\mathrm{H}>\SI{3.9e24}{\per\centi\meter\squared}$ to produce the observed {2-10\,\si{\kilo\electronvolt}} luminosity. 
Increasing the photon index to $\Gamma=2$ would raise the column density to $N_\mathrm{H}>\SI{4.7e24}{\per\centi\meter\squared}$.
In other words, the AGN needs to be Compton-thick along the line of sight to explain the non-detection of X-rays.
However, at the same time, there must be an optically thin path for the ionizing radiation to escape and produce the observed \oiii emission.

We repeated the exercise using the \singlet{Ne}{v}{3427} detection. Given the extremely high energies needed to produce the Ne$^{4+}$ ion, the contribution from SF is null or negligible. In the O3-N-core aperture, we measured a \singlet{Ne}{v}{3427} flux of \SI{2.25+-0.33e-18}{\erg\per\second\per\centi\meter\squared} and a Balmer decrement of $\mathrm{H}\alpha/\mathrm{H}\beta=\num{4.52+-0.16}$. Assuming case B recombination with an intrinsic ratio of $\mathrm{H}\alpha/\mathrm{H}\beta=\num{2.86}$ and a \citep{Calzetti2000Dust} attenuation law, we derive $A_V=\num{1.58+-0.16}$ mag. We thus inferred a reddening-corrected \singlet{Ne}{v}{3427} luminosity of \SI{4.8+-1.2e42}{\erg\per\second}.
\citet{Berney2015BassXrayOpticalLines} also provide a relation for [Ne\,\textsc{v}] luminosity versus X-ray luminosity, although it is derived from a smaller sample and has a larger scatter than the \oiii relation. According to this latter relation, the intrinsic X-ray luminosity should be \SI{2.4e45}{\erg\per\second}. 
%Compared to the prediction based on \oiii, one could argue that the AGN is responsible for $\sim 1/4$ of the \oiii emission.
Using the same model as before, the column density needed to obscure that X-ray output is $N_\mathrm{H}>\SI{2e24}{\per\centi\meter\squared}$, which is just above the Compton-thick limit.

Assuming the AGN is buried at the center of the DSFG, we now ask how much of the inferred obscuration can be accounted for by the ISM alone. This is motivated by the recent results of \citet{Andonie2024ObscurationBeyondTheNucleus} who find that infrared-quasar host galaxies with SFR$\gtrsim\SI{300}{\msun\per\year}$ have submillimeter sizes as compact as those of DSFGs, implying a very dense ISM with column densities potentially exceeding the Compton limit. Following \citet{Andonie2024ObscurationBeyondTheNucleus}, we estimated the average column density by uniformly distributing the total gas mass (here $M_\mathrm{gas}=\SI{1.5e11}{\msun}$, \citealt{Fraternali2021FastRotators}) over a sphere with a radius equal to the S\'ersic effective radius ($r_\mathrm{eff}=\SI{0.74}{\kilo\parsec}$, measured from a fit to the rest-frame \SI{158}{\micro\meter} ALMA image, \citealt{Solimano2024CristalPlume}), finding $\left<N_\mathrm{H}\right>_{\mathrm{ISM}}\approx \SI{2.5e24}{\per\centi\meter\squared}$. Therefore, we conclude that the ISM of the DSFG has enough material to obscure the X-rays. However, some of it must be located very close to the AGN; otherwise, we would detect the broad line region in the spectrum of the DSFG.

The results discussed above again suggest that the AGN is heavily obscured toward our line of sight, yet is powerful enough to produce luminous \oiii (and even [Ne\,\textsc{v}] emission) along unobscured sightlines. Such sightlines can be the result of a radio jet that has cleared them of obscuring material;  we discuss this explanation in Sect.~\ref{sec:scenario}.

%The main caveat is that this conclusion relies on \singlet{Ne}{v}{3427} being produced entirely by photoionization, rather than shocks.
\subsection{Proposed scenario and implications}\label{sec:scenario}

Here, we put forward a scenario that explains the observed emission.
First, we assume that an AGN resides in the very center of the DSFG. 
This is motivated by the fact that the DSFG occupies the AGN loci of all the diagnostic diagrams we have considered (see Sec.~\ref{sec:results:bpt}). 
In addition, given the M$_\mathrm{SMBH}-\mathrm{M}_{*}$ relation \citep[e.g.,][]{ReinesAndVolonteri2015BhMassVsMstar,Pacucci2023CeersJadesAgn}, the DSFG is the most likely to host a massive SMBH, and thus an AGN. 
Moreover, its location between the two radio detections makes it the potential launching site of a jet, as proposed by \citet{Jimenez-Andrade2023LyaNebulaeGalaxyPair}.

An alternative explanation has O3-N as a separate galaxy altogether; presumably a lower-mass AGN host currently being accreted by the DSFG (see scenarios 3 and 4 of \citealt{Solimano2024CristalPlume}). This would mean that the \sysname system hosts either a dual AGN (if both the DSFG and O3-N are active) or an offset AGN (if only O3-N is active). Dual and offset AGN have been extensively reported in the literature, with examples in both the local \citep[e.g.,][]{Barth2008OffsetAgnNgc3341, Mazzarella2012DualAgnMrk266, Koss2012UnderstandingDualAgn, Barrows2017OffsetAgnTriggering, Secrest2017Was49bLocalOffsetAgn} and distant Universe \citep[e.g.,][]{Gerke2007Deep2DualAgnz0p7, Comerford2015MergerDrivenAgn, Perna2023GaNifsDualAgnEnvironment, Ubler2024OffsetAgnGaNifs}. These phenomena are strongly linked to galaxy mergers, making \sysname a plausible candidate. However, an investigation of the 3D geometry of the system to decipher whether O3-N is behind or in front of the DSFG is required to provide further insight into the offset/dual AGN scenario. We defer such an analysis to a future paper.
%An alternative situation is that O3-N is a separate galaxy altogether, presumably a lower mass AGN host currently being accreted by the DSFG. \comment{expand on this idea? I feel I don't have evidence against this. Also, connect to the scenarios described in Solimano24a}

For simplicity, we return to our fiducial scenario where the AGN is at the nucleus of the DSFG. In this picture, O3-N traces an extended emission-line region \citep*[EELR; e.g., ][]{Stockton2006QsoExtendedEmissionLineRegions} and an outflow driven by the AGN, as evidenced by the line ratios and broad velocity component, respectively.

Regarding O3-S, one could presume it represents the bipolar counterpart of O3-N (i.e., the receding side of the outflow). However, the observed kinematics, morphology, and spectral properties suggest otherwise.
In particular, the large velocity gradient, narrower line width, lower SB, lower metallicity (see Sect.~\ref{sec:metallicity}), and more elongated structure make O3-S fundamentally different from O3-N.
As suggested in Sect.~\ref{sec:results:morphokin}, O3-S is unlikely to be a separate galaxy with M$_\mathrm{dyn}\approx\SI{e11}{\msun}$, but rather a stream of tidal debris.
The connection (both spatial and spectral) between C01 and O3-S then suggests that the gas might have been tidally stripped from C01. This is supported by the finding that O3-S has a very similar oxygen abundance to the southern clump of C01 (C01-SW, see Fig.~\ref{fig:metallicities}).

Finally, due to the high ionization implied by the strong \oiii and He\,\textsc{ii} lines, and its location along the jet axis, we propose that O3-S is being externally illuminated by the AGN. 
In other words, O3-S is an EELR analog to the famous Hanny's Voorwerp \citep{Lintott2009HannyVoorwerp}. 
The Voorwerp is characterized by extended, high-equivalent-width \oiii emission at a far projected distance from the galaxy IC\,2497.
The leading explanation for the nature of the Voorwerp is that a portion of an otherwise-invisible gas tidal tail was exposed to ionizing radiation from the now-faded AGN in the center of IC\,2497.
Moreover, the escape path of the ionizing photons was carved by a past jet, as evidenced by the detection of a steep-spectrum extended radio relic \citep{Josza2009RadioObsHsV, Smith2022RadioRelicHsVLofar}.

It could be that we are witnessing a similar situation in \sysname, although we cannot say whether the AGN in the DSFG is currently switched off or is simply obscured along the line of sight. 
However, the radio luminosity in \sysname is approximately 100 times lower than the power of typical radio-selected HzRGs \citep[$L_{1.4\,\mathrm{GHz}}\gtrsim \SI{e27}{\watt\per\hertz}$,][]{MileyAndDeBreuck2008HzrgsAndEnvironments}, but at the same time is approximately 500 times higher than that of the Voorwerp \citep[$L_{1.4\,\mathrm{GHz}}\approx \SI{e23}{\watt\per\hertz}$,][]{Josza2009RadioObsHsV}. 
Therefore,  the observed emission is more likely explained by a moderate luminosity jet, rather than a relic. 
Therefore, when considering the alignment of the radio sources along the axis that connects O3-N and O3-S, the idea of a jet carving an unobscured sightline in the polar direction becomes more compelling.

It is important to emphasize that \sysname was selected because of its bright submillimeter emission and associated \lya blob, as this could uncover a common trend among DSFG groups or protoclusters. For example, \citet{Vito2020ComptonThickQsoInDrc} find a powerful ($L_{2-10\,\mathrm{keV}}\approx\SI{3e45}{\erg\per\second}$) Compton-thick QSO hiding in the most gas-rich and submillimeter-bright member of the $z=4$ protocluster known as DRC \citep{Oteo2018DRC}. This system also hosts a LAB with emission of the high-ionization \heii and \civ lines. A second example is that of the SPT2349-56 protocluster, a $z=4.3$ structure hosting at least 21 DSFGs \citep{Miller2018SPT2349,Hill2020SPT2349} and a faint LAB \citep{Apostolovski2024ExtendedLyaSpt2349}. Recently, deep X-ray and radio observations revealed the presence of two AGN within SPT2349-56 \citep{Vito2024XraySmbhSpt2349, Chapman2024RadioAgnSpt2349}.
\citet{Vito2024XraySmbhSpt2349} then argue that SPT2349-56 and DRC together provide evidence of an enhanced AGN fraction within gas-rich protoclusters.

\sysname is certainly less massive and extreme than either DRC or SPT2349-56, yet our detection of a Compton-thick AGN in it may be related to the same mechanisms that trigger obscured SMBH accretion in these two protoclusters. In theoretical frameworks of galaxy--SMBH coevolution \citep[e.g., ][]{Hopkins2008CosmoFramework1Mergers}, it is expected that the conditions (e.g., those created in gas-rich mergers) that give rise to extreme dusty starbursts  also favor (obscured) SMBH accretion. Observationally, the DSFG-AGN connection is stronger in dense environments, which is possibly due to the higher rate of merger and interaction \citep{Monson2023XraysSsa22ProtoCluster, Vito2024XraySmbhSpt2349}. A more systematic investigation of the obscured AGN fraction as a function of overdensity and SFR is needed to consolidate these trends.

While rest-frame hard X-rays will remain the gold standard for selecting obscured AGN, current facilities need to spend several tens of hours on source to produce detections at high $z$. In this context, our results suggest that a search for EELRs with the JWST NIRCam and NIRSpec might provide an alternative way to select high-$z$ AGN.

\section{Summary and conclusions}\label{sec:conclusions}

We present the discovery and characterization of two bright \oiii nebulae, O3-N and O3-S, around the \sysname DSFG. 
Using JWST/NIRCam and JWST/NIRSpec, we characterized the morpho-kinematic structure of the nebulae, as well as their potential sources of ionization.
Our results can be summarized as follows:
\begin{itemize}
    \item O3-N, the brightest \oiii nebula in the system, shows a broad and blueshifted velocity component with FWHM$\gtrsim\SI{1200}{\kilo\meter\per\second}$, as measured in both \oiii and \halpha lines.
	    We interpret this as evidence of ionized outflows, with a potential link to the \cii plume of \citet{Solimano2024CristalPlume}.
    
    \item While fainter than O3-N, O3-S is more extended and shows an elongated but irregular morphology.
    Moreover, the resolved \oiii velocity field reveals a \SI{800}{\kilo\meter\per\second} gradient roughly aligned with the major axis of O3-S, but without a peaked velocity dispersion profile. 
    Also, the lack of emission from stars, cold gas or dust from O3-S disfavors its identification as a massive rotating galaxy.
    Instead, given the low-SB bridge between O3-S and C01, in addition to their similar oxygen abundances, we deem it more likely that O3-S is a tidal feature stemming from C01.
    
    \item Nebular line ratio diagrams suggest at least some degree of AGN ionization in all the sources considered in this paper (except for C01).
	    The DSFG, in particular, shows line ratios consistent with AGN in all the diagrams considered. 
    \item We detect the temperature-sensitive \singlet{O}{iii}{4634} auroral line in the central regions of O3-N and O3-S. However, these sources fall very close to the boundary line between pure AGN and an AGN--SF mixture in the \singlet{O}{iii}{4634}-based diagrams proposed by \citet{Mazzolari24NewAgnDiagnostic}.
    
    \item The central region of O3-N also shows a significant detection of the high-ionization \singlet{Ne}{v}{3427} line ($E>\SI{97.1}{\electronvolt}$), an almost univocal tracer of AGN activity. 
	    Paired with the non-detection of rest-frame hard X-rays, we derive low $L_{X}/L_{[\mathrm{O}\, \mathrm{III}]}$ and $L_{X}/L_{[\mathrm{Ne}\,\mathrm{V}]}$ ratios that imply Compton-thick levels of obscuration. 
	    Interestingly, if the AGN is at the nucleus of the DSFG, the inferred column densities ($N_\mathrm{H}\gtrsim2-\SI{5e24}{\per\centi\meter\squared}$) are consistent with 
	    arising from the ISM alone, as suggested by its dense and compact morphology, as derived in previous ALMA imaging.

    \item We tested whether or not shocks could be responsible for the emission in O3-N and O3-S by measuring the \singlet{N}{ii}{6850}/\halpha ratio on a spaxel-by-spaxel basis. We find no correlation between the ratio and the velocity dispersion, thus disfavoring a shock scenario.
\end{itemize}

We propose a scenario where both nebulae are EELRs powered by an AGN deeply buried within the DSFG.
While O3-N shows a prominent outflow, O3-S belongs to a tidal tail of C01.
This scenario makes O3-S a plausible high-$z$ analog of Hanny's Voorwerp, a residual ionized nebula excited by a faded AGN. In this picture, the action of the radio jets might have opened a path along the polar direction for the AGN ionizing radiation to reach O3-S and O3-N.
We discuss our findings in the context of the enhanced AGN fraction found in massive, gas-rich protocluster cores at similar redshifts. Our results suggest that the processes that drive obscured SMBH accretion in these structures (e.g., mergers) might also be at play in the less massive \sysname group.
Finally, we highlight the ability of JWST to uncover hidden AGN at high redshifts, in a regime where current X-ray facilities lack the required sensitivity.

\begin{acknowledgements}
This work is based in part on observations made with the NASA/ESA/CSA James Webb Space Telescope.
The data were obtained from the Mikulski Archive for Space Telescopes at the Space Telescope Science Institute, which is operated by the Association of Universities for Research in Astronomy, Inc., under NASA contract NAS 5-03127 for JWST.
These observations are associated with programs JWST-GO-01727, JWST-GO-0345, and JWST-GO-04265. We also thank Mingyu Li for useful discussions and the My Filter tool \href{https://doi.org/10.5281/zenodo.10210201}{10.5281/zenodo.10210201}.
M. S. was financially supported by Becas-ANID scholarship \#21221511. M. S., S. B., M. A., R. J. A., J. G-L., M. Boquien, and V. V. all acknowledge support from ANID BASAL project FB210003. 
M. R. acknowledges support from the Narodowe Centrum Nauki (UMO-2020/38/E/ST9/00077) and support from the Foundation for Polish Science (FNP) under the program START 063.2023.
E. I. acknowledges funding by ANID FONDECYT Regular 1221846.
M. Boquien gratefully acknowledges support  from the FONDECYT regular grant 1211000.
This work was supported by the French government through the France 2030 investment plan managed by the National Research Agency (ANR), as part of the Initiative of Excellence of Universit\'e C{\^o}te d’Azur under reference number ANR-15-IDEX-01.
G. C. J. acknowledges funding from the ``FirstGalaxies'' Advanced Grant from the European Research Council (ERC) under the European Union’s Horizon 2020 research and innovation programme (Grant agreement No. 78905).
R. J. A. was supported by FONDECYT grant number 1231718.
 H.I. acknowledges support from JSPS KAKENHI Grant Number JP21H01129 and the Ito Foundation for Promotion of Science.
R. L. D is supported by the Australian Research Council through the Discovery Early Career Researcher Award (DECRA) Fellowship DE240100136 funded by the Australian Government. 
\end{acknowledgements}

\bibliographystyle{aa}
\bibliography{main.bib}

\begin{thebibliography}{107}
\expandafter\ifx\csname natexlab\endcsname\relax\def\natexlab#1{#1}\fi

\bibitem[{{Alarie} \& {Morisset}(2019)}]{AlarieMorisset2019Mappings5ShockDb}
{Alarie}, A. \& {Morisset}, C. 2019, \rmxaa, 55, 377

\bibitem[{{Allen} {et~al.}(2008){Allen}, {Groves}, {Dopita}, {Sutherland}, \& {Kewley}}]{Allen2008Mappings3ShockLibrary}
{Allen}, M.~G., {Groves}, B.~A., {Dopita}, M.~A., {Sutherland}, R.~S., \& {Kewley}, L.~J. 2008, \apjs, 178, 20

\bibitem[{{Andonie} {et~al.}(2024){Andonie}, {Alexander}, {Greenwell}, {Puglisi}, {Laloux}, {Alonso-Tetilla}, {Calistro Rivera}, {Harrison}, {Hickox}, {Kaasinen}, {Lapi}, {L{\'o}pez}, {Petter}, {Ramos Almeida}, {Rosario}, {Shankar}, \& {Villforth}}]{Andonie2024ObscurationBeyondTheNucleus}
{Andonie}, C., {Alexander}, D.~M., {Greenwell}, C., {et~al.} 2024, \mnras, 527, L144

\bibitem[{{Apostolovski} {et~al.}(2024){Apostolovski}, {Aravena}, {Anguita}, {Bethermin}, {Burgoyne}, {Chapman}, {De Breuck}, {Gonzalez}, {Gronke}, {Guaita}, {Hezaveh}, {Hill}, {Jarugula}, {Johnston}, {Malkan}, {Narayanan}, {Reuter}, {Solimano}, {Spilker}, {Sulzenauer}, {Vieira}, {Vizgan}, \& {Wei{\ss}}}]{Apostolovski2024ExtendedLyaSpt2349}
{Apostolovski}, Y., {Aravena}, M., {Anguita}, T., {et~al.} 2024, \aap, 683, A64

\bibitem[{{Bagley} {et~al.}(2023){Bagley}, {Finkelstein}, {Koekemoer}, {Ferguson}, {Arrabal Haro}, {Dickinson}, {Kartaltepe}, {Papovich}, {P{\'e}rez-Gonz{\'a}lez}, {Pirzkal}, {Somerville}, {Willmer}, {Yang}, {Yung}, {Fontana}, {Grazian}, {Grogin}, {Hirschmann}, {Kewley}, {Kirkpatrick}, {Kocevski}, {Lotz}, {Medrano}, {Morales}, {Pentericci}, {Ravindranath}, {Trump}, {Wilkins}, {Calabr{\`o}}, {Cooper}, {Costantin}, {de la Vega}, {Hilbert}, {Hutchison}, {Larson}, {Lucas}, {McGrath}, {Ryan}, {Wang}, \& {Wuyts}}]{Bagley2023CeersNircamReduction}
{Bagley}, M.~B., {Finkelstein}, S.~L., {Koekemoer}, A.~M., {et~al.} 2023, \apjl, 946, L12

\bibitem[{{Baldwin} {et~al.}(1981){Baldwin}, {Phillips}, \& {Terlevich}}]{Baldwin1981BPTpaper}
{Baldwin}, J.~A., {Phillips}, M.~M., \& {Terlevich}, R. 1981, \pasp, 93, 5

\bibitem[{{Barrows} {et~al.}(2017){Barrows}, {Comerford}, {Greene}, \& {Pooley}}]{Barrows2017OffsetAgnTriggering}
{Barrows}, R.~S., {Comerford}, J.~M., {Greene}, J.~E., \& {Pooley}, D. 2017, \apj, 838, 129

\bibitem[{{Barth} {et~al.}(2008){Barth}, {Bentz}, {Greene}, \& {Ho}}]{Barth2008OffsetAgnNgc3341}
{Barth}, A.~J., {Bentz}, M.~C., {Greene}, J.~E., \& {Ho}, L.~C. 2008, \apjl, 683, L119

\bibitem[{{Baugh} {et~al.}(1998){Baugh}, {Cole}, {Frenk}, \& {Lacey}}]{Baugh1998EpochOfGalaxyFormation}
{Baugh}, C.~M., {Cole}, S., {Frenk}, C.~S., \& {Lacey}, C.~G. 1998, \apj, 498, 504

\bibitem[{{Berney} {et~al.}(2015){Berney}, {Koss}, {Trakhtenbrot}, {Ricci}, {Lamperti}, {Schawinski}, {Balokovi{\'c}}, {Crenshaw}, {Fischer}, {Gehrels}, {Harrison}, {Hashimoto}, {Ichikawa}, {Mushotzky}, {Oh}, {Stern}, {Treister}, {Ueda}, {Veilleux}, \& {Winter}}]{Berney2015BassXrayOpticalLines}
{Berney}, S., {Koss}, M., {Trakhtenbrot}, B., {et~al.} 2015, \mnras, 454, 3622

\bibitem[{{Borisova} {et~al.}(2016){Borisova}, {Cantalupo}, {Lilly}, {Marino}, {Gallego}, {Bacon}, {Blaizot}, {Bouch{\'e}}, {Brinchmann}, {Carollo}, {Caruana}, {Finley}, {Herenz}, {Richard}, {Schaye}, {Straka}, {Turner}, {Urrutia}, {Verhamme}, \& {Wisotzki}}]{Borisova2016MUSELyaBlobsAroundQSOs}
{Borisova}, E., {Cantalupo}, S., {Lilly}, S.~J., {et~al.} 2016, \apj, 831, 39

\bibitem[{{Cai} {et~al.}(2017){Cai}, {Fan}, {Yang}, {Bian}, {Prochaska}, {Zabludoff}, {McGreer}, {Zheng}, {Green}, {Cantalupo}, {Frye}, {Hamden}, {Jiang}, {Kashikawa}, \& {Wang}}]{Cai2017ElaneInOverdensity}
{Cai}, Z., {Fan}, X., {Yang}, Y., {et~al.} 2017, \apj, 837, 71

\bibitem[{{Calzetti} {et~al.}(2000){Calzetti}, {Armus}, {Bohlin}, {Kinney}, {Koornneef}, \& {Storchi-Bergmann}}]{Calzetti2000Dust}
{Calzetti}, D., {Armus}, L., {Bohlin}, R.~C., {et~al.} 2000, \apj, 533, 682

\bibitem[{{Capak} {et~al.}(2008){Capak}, {Carilli}, {Lee}, {Aldcroft}, {Aussel}, {Schinnerer}, {Wilson}, {Yun}, {Blain}, {Giavalisco}, {Ilbert}, {Kartaltepe}, {Lee}, {McCracken}, {Mobasher}, {Salvato}, {Sasaki}, {Scott}, {Sheth}, {Shioya}, {Thompson}, {Elvis}, {Sanders}, {Scoville}, \& {Tanaguchi}}]{Capak2008ExtremeStarburst}
{Capak}, P., {Carilli}, C.~L., {Lee}, N., {et~al.} 2008, \apjl, 681, L53

\bibitem[{{Cappellari}(2017)}]{Cappellari2017FullSpectrumFitting}
{Cappellari}, M. 2017, \mnras, 466, 798

\bibitem[{{Cappellari}(2023)}]{Cappellari2023PpxfUpdate}
{Cappellari}, M. 2023, \mnras, 526, 3273

\bibitem[{{Carilli} {et~al.}(2008){Carilli}, {Lee}, {Capak}, {Schinnerer}, {Lee}, {McCraken}, {Yun}, {Scoville}, {Smol{\v{c}}i{\'c}}, {Giavalisco}, {Datta}, {Taniguchi}, \& {Urry}}]{Carilli2008RadioLbgs}
{Carilli}, C.~L., {Lee}, N., {Capak}, P., {et~al.} 2008, \apj, 689, 883

\bibitem[{{Casey}(2016)}]{Casey2016UbiquityOfCoevalStarbursts}
{Casey}, C.~M. 2016, \apj, 824, 36

\bibitem[{{Casey} {et~al.}(2023){Casey}, {Kartaltepe}, {Drakos}, {Franco}, {Harish}, {Paquereau}, {Ilbert}, {Rose}, {Cox}, {Nightingale}, {Robertson}, {Silverman}, {Koekemoer}, {Massey}, {McCracken}, {Rhodes}, {Akins}, {Allen}, {Amvrosiadis}, {Arango-Toro}, {Bagley}, {Bongiorno}, {Capak}, {Champagne}, {Chartab}, {Ch{\'a}vez Ortiz}, {Chworowsky}, {Cooke}, {Cooper}, {Darvish}, {Ding}, {Faisst}, {Finkelstein}, {Fujimoto}, {Gentile}, {Gillman}, {Gould}, {Gozaliasl}, {Hayward}, {He}, {Hemmati}, {Hirschmann}, {Jahnke}, {Jin}, {Khostovan}, {Kokorev}, {Lambrides}, {Laigle}, {Larson}, {Leung}, {Liu}, {Liaudat}, {Long}, {Magdis}, {Mahler}, {Mainieri}, {Manning}, {Maraston}, {Martin}, {McCleary}, {McKinney}, {McPartland}, {Mobasher}, {Pattnaik}, {Renzini}, {Rich}, {Sanders}, {Sattari}, {Scognamiglio}, {Scoville}, {Sheth}, {Shuntov}, {Sparre}, {Suzuki}, {Talia}, {Toft}, {Trakhtenbrot}, {Urry}, {Valentino}, {Vanderhoof}, {Vardoulaki}, {Weaver}, {Whitaker}, {Wilkins}, {Yang}, \& {Zavala}}]{Casey2023CosmosWebOverview}
{Casey}, C.~M., {Kartaltepe}, J.~S., {Drakos}, N.~E., {et~al.} 2023, \apj, 954, 31

\bibitem[{Cavanaugh(1997)}]{Cavanaugh1997CorrectedAkaikeCriterion}
Cavanaugh, J.~E. 1997, Statistics \& Probability Letters, 33, 201

\bibitem[{{Chapman} {et~al.}(2024){Chapman}, {Hill}, {Aravena}, {Archipley}, {Babul}, {Burgoyne}, {Canning}, {Deane}, {De Breuck}, {Gonzalez}, {Hayward}, {Kim}, {Malkan}, {Marrone}, {McIntyre}, {Murphy}, {Pass}, {Perry}, {Phadke}, {Rennehan}, {Reuter}, {Rotermund}, {Scott}, {Seymour}, {Solimano}, {Spilker}, {Stark}, {Sulzenauer}, {Tothill}, {Vieira}, {Vizgan}, {Wang}, \& {Weiss}}]{Chapman2024RadioAgnSpt2349}
{Chapman}, S.~C., {Hill}, R., {Aravena}, M., {et~al.} 2024, \apj, 961, 120

\bibitem[{{Cleri} {et~al.}(2023){Cleri}, {Olivier}, {Hutchison}, {Papovich}, {Trump}, {Amor{\'\i}n}, {Backhaus}, {Berg}, {Fern{\'a}ndez}, {Finkelstein}, {Fujimoto}, {Hirschmann}, {Kartaltepe}, {Kocevski}, {Simons}, {Wilkins}, \& {Yung}}]{Cleri2023Ne53RatioEig}
{Cleri}, N.~J., {Olivier}, G.~M., {Hutchison}, T.~A., {et~al.} 2023, \apj, 953, 10

\bibitem[{{Comerford} {et~al.}(2015){Comerford}, {Pooley}, {Barrows}, {Greene}, {Zakamska}, {Madejski}, \& {Cooper}}]{Comerford2015MergerDrivenAgn}
{Comerford}, J.~M., {Pooley}, D., {Barrows}, R.~S., {et~al.} 2015, \apj, 806, 219

\bibitem[{{Conroy} \& {Gunn}(2010)}]{Conroy2010FspsCalib}
{Conroy}, C. \& {Gunn}, J.~E. 2010, \apj, 712, 833

\bibitem[{{Conroy} {et~al.}(2009){Conroy}, {Gunn}, \& {White}}]{Conroy2009FspsImf}
{Conroy}, C., {Gunn}, J.~E., \& {White}, M. 2009, \apj, 699, 486

\bibitem[{{Decarli} {et~al.}(2024){Decarli}, {Loiacono}, {Farina}, {Dotti}, {Lupi}, {Meyer}, {Mignoli}, {Pensabene}, {Strauss}, {Venemans}, {Yang}, {Walter}, {Wolf}, {Ba{\~n}ados}, {Blecha}, {Bosman}, {Carilli}, {Comastri}, {Connor}, {Costa}, {Eilers}, {Fan}, {Gilli}, {Jun}, {Liu}, {Marshall}, {Mazzucchelli}, {Neeleman}, {Onoue}, {Overzier}, {Pudoka}, {Riechers}, {Rix}, {Schindler}, {Trakhtenbrot}, {Trebitsch}, {Vestergaard}, {Volonteri}, {Wang}, {Zhang}, \& {Zou}}]{Decarli2024NIrspecResdshiftSixQso}
{Decarli}, R., {Loiacono}, F., {Farina}, E.~P., {et~al.} 2024, \aap, 689, A219

\bibitem[{{Decarli} {et~al.}(2019){Decarli}, {Mignoli}, {Gilli}, {Balmaverde}, {Brusa}, {Cappelluti}, {Comastri}, {Nanni}, {Peca}, {Pensabene}, {Vanzella}, \& {Vignali}}]{Decarli2019FirstHighZQsoOverdensity}
{Decarli}, R., {Mignoli}, M., {Gilli}, R., {et~al.} 2019, \aap, 631, L10

\bibitem[{{D'Eugenio} {et~al.}(2024){D'Eugenio}, {Maiolino}, {Mahatma}, {Mazzolari}, {Carniani}, {de Graaff}, {Maseda}, {Parlanti}, {Bunker}, {Ji}, {Jones}, {Morganti}, {Scholtz}, {Tacchella}, {Tadhunter}, {{\"U}bler}, \& {Venturi}}]{DEugenio2024NirspecWideShockAgn}
{D'Eugenio}, F., {Maiolino}, R., {Mahatma}, V.~H., {et~al.} 2024, arXiv e-prints, arXiv:2408.03982

\bibitem[{{Dopita} {et~al.}(2016){Dopita}, {Kewley}, {Sutherland}, \& {Nicholls}}]{Dopita2016MetallicityIndicator}
{Dopita}, M.~A., {Kewley}, L.~J., {Sutherland}, R.~S., \& {Nicholls}, D.~C. 2016, \apss, 361, 61

\bibitem[{{Elvis} {et~al.}(2009){Elvis}, {Civano}, {Vignali}, {Puccetti}, {Fiore}, {Cappelluti}, {Aldcroft}, {Fruscione}, {Zamorani}, {Comastri}, {Brusa}, {Gilli}, {Miyaji}, {Damiani}, {Koekemoer}, {Finoguenov}, {Brunner}, {Urry}, {Silverman}, {Mainieri}, {Hasinger}, {Griffiths}, {Carollo}, {Hao}, {Guzzo}, {Blain}, {Calzetti}, {Carilli}, {Capak}, {Ettori}, {Fabbiano}, {Impey}, {Lilly}, {Mobasher}, {Rich}, {Salvato}, {Sanders}, {Schinnerer}, {Scoville}, {Shopbell}, {Taylor}, {Taniguchi}, \& {Volonteri}}]{Elvis2009ChandraCosmosOverview}
{Elvis}, M., {Civano}, F., {Vignali}, C., {et~al.} 2009, \apjs, 184, 158

\bibitem[{{Emonts} {et~al.}(2018){Emonts}, {Lehnert}, {Dannerbauer}, {De Breuck}, {Villar-Mart{\'\i}n}, {Miley}, {Allison}, {Gullberg}, {Hatch}, {Guillard}, {Mao}, \& {Norris}}]{Emonts2018SpiderwebCircumgalacticCi}
{Emonts}, B.~H.~C., {Lehnert}, M.~D., {Dannerbauer}, H., {et~al.} 2018, \mnras, 477, L60

\bibitem[{{Emonts} {et~al.}(2023){Emonts}, {Lehnert}, {Yoon}, {Mandelker}, {Villar-Mart{\'\i}n}, {Miley}, {De Breuck}, {P{\'e}rez-Torres}, {Hatch}, \& {Guillard}}]{Emonts2023CarbonCosmicStream}
{Emonts}, B. H.~C., {Lehnert}, M.~D., {Yoon}, I., {et~al.} 2023, Science, 379, 1323

\bibitem[{{Fraternali} {et~al.}(2021){Fraternali}, {Karim}, {Magnelli}, {G{\'o}mez-Guijarro}, {Jim{\'e}nez-Andrade}, \& {Posses}}]{Fraternali2021FastRotators}
{Fraternali}, F., {Karim}, A., {Magnelli}, B., {et~al.} 2021, \aap, 647, A194

\bibitem[{{Gerke} {et~al.}(2007){Gerke}, {Newman}, {Lotz}, {Yan}, {Barmby}, {Coil}, {Conselice}, {Ivison}, {Lin}, {Koo}, {Nandra}, {Salim}, {Small}, {Weiner}, {Cooper}, {Davis}, {Faber}, \& {Guhathakurta}}]{Gerke2007Deep2DualAgnz0p7}
{Gerke}, B.~F., {Newman}, J.~A., {Lotz}, J., {et~al.} 2007, \apjl, 660, L23

\bibitem[{{Gilli} {et~al.}(2010){Gilli}, {Vignali}, {Mignoli}, {Iwasawa}, {Comastri}, \& {Zamorani}}]{Gilli2010XrayToNevRatio}
{Gilli}, R., {Vignali}, C., {Mignoli}, M., {et~al.} 2010, \aap, 519, A92

\bibitem[{{G{\'o}mez-Guijarro} {et~al.}(2018){G{\'o}mez-Guijarro}, {Toft}, {Karim}, {Magnelli}, {Magdis}, {Jim{\'e}nez-Andrade}, {Capak}, {Fraternali}, {Fujimoto}, {Riechers}, {Schinnerer}, {Smol{\v{c}}i{\'c}}, {Aravena}, {Bertoldi}, {Cortzen}, {Hasinger}, {Hu}, {Jones}, {Koekemoer}, {Lee}, {McCracken}, {Micha{\l}owski}, {Navarrete}, {Povi{\'c}}, {Puglisi}, {Romano-D{\'\i}az}, {Sheth}, {Silverman}, {Staguhn}, {Steinhardt}, {Stockmann}, {Tanaka}, {Valentino}, {van Kampen}, \& {Zirm}}]{GomezGuijarro2018AlmaMinorMergersSmgs}
{G{\'o}mez-Guijarro}, C., {Toft}, S., {Karim}, A., {et~al.} 2018, \apj, 856, 121

\bibitem[{{Guaita} {et~al.}(2022){Guaita}, {Aravena}, {Gurung-Lopez}, {Cantalupo}, {Marino}, {Riechers}, {da Cunha}, {Wagg}, {Algera}, {Dannerbauer}, \& {Cox}}]{Guaita2022AzTec3Lya}
{Guaita}, L., {Aravena}, M., {Gurung-Lopez}, S., {et~al.} 2022, \aap, 660, A137

\bibitem[{{Harikane} {et~al.}(2023){Harikane}, {Zhang}, {Nakajima}, {Ouchi}, {Isobe}, {Ono}, {Hatano}, {Xu}, \& {Umeda}}]{Harikane2023NirspecFirstAgnCensus}
{Harikane}, Y., {Zhang}, Y., {Nakajima}, K., {et~al.} 2023, \apj, 959, 39

\bibitem[{{Hennawi} {et~al.}(2015){Hennawi}, {Prochaska}, {Cantalupo}, \& {Arrigoni-Battaia}}]{Hennawi2015QuasarGiantLyaNebula}
{Hennawi}, J.~F., {Prochaska}, J.~X., {Cantalupo}, S., \& {Arrigoni-Battaia}, F. 2015, Science, 348, 779

\bibitem[{{Hill} {et~al.}(2020){Hill}, {Chapman}, {Scott}, {Apostolovski}, {Aravena}, {B{\'e}thermin}, {Bradford}, {Canning}, {De Breuck}, {Dong}, {Gonzalez}, {Greve}, {Hayward}, {Hezaveh}, {Litke}, {Malkan}, {Marrone}, {Phadke}, {Reuter}, {Rotermund}, {Spilker}, {Vieira}, \& {Wei{\ss}}}]{Hill2020SPT2349}
{Hill}, R., {Chapman}, S., {Scott}, D., {et~al.} 2020, \mnras, 495, 3124

\bibitem[{{Ho} {et~al.}(2014){Ho}, {Kewley}, {Dopita}, {Medling}, {Allen}, {Bland-Hawthorn}, {Bloom}, {Bryant}, {Croom}, {Fogarty}, {Goodwin}, {Green}, {Konstantopoulos}, {Lawrence}, {L{\'o}pez-S{\'a}nchez}, {Owers}, {Richards}, \& {Sharp}}]{Ho2014SamiShocksAndOutflows}
{Ho}, I.~T., {Kewley}, L.~J., {Dopita}, M.~A., {et~al.} 2014, \mnras, 444, 3894

\bibitem[{{Hopkins} {et~al.}(2008){Hopkins}, {Hernquist}, {Cox}, \& {Kere{\v{s}}}}]{Hopkins2008CosmoFramework1Mergers}
{Hopkins}, P.~F., {Hernquist}, L., {Cox}, T.~J., \& {Kere{\v{s}}}, D. 2008, \apjs, 175, 356

\bibitem[{{Izotov} {et~al.}(2012){Izotov}, {Thuan}, \& {Privon}}]{Izotov2012NevEmissionBlueCompactDwarfs}
{Izotov}, Y.~I., {Thuan}, T.~X., \& {Privon}, G. 2012, \mnras, 427, 1229

\bibitem[{{Jim{\'e}nez-Andrade} {et~al.}(2023){Jim{\'e}nez-Andrade}, {Cantalupo}, {Magnelli}, {Romano-D{\'\i}az}, {G{\'o}mez-Guijarro}, {Mackenzie}, {Smol{\v{c}}i{\'c}}, {Murphy}, {Matthee}, \& {Toft}}]{Jimenez-Andrade2023LyaNebulaeGalaxyPair}
{Jim{\'e}nez-Andrade}, E.~F., {Cantalupo}, S., {Magnelli}, B., {et~al.} 2023, \mnras, 521, 2326

\bibitem[{{J{\'o}zsa} {et~al.}(2009){J{\'o}zsa}, {Garrett}, {Oosterloo}, {Rampadarath}, {Paragi}, {van Arkel}, {Lintott}, {Keel}, {Schawinski}, \& {Edmondson}}]{Josza2009RadioObsHsV}
{J{\'o}zsa}, G.~I.~G., {Garrett}, M.~A., {Oosterloo}, T.~A., {et~al.} 2009, \aap, 500, L33

\bibitem[{{Kewley} {et~al.}(2001){Kewley}, {Dopita}, {Sutherland}, {Heisler}, \& {Trevena}}]{Kewley2001ModelingStarbursts}
{Kewley}, L.~J., {Dopita}, M.~A., {Sutherland}, R.~S., {Heisler}, C.~A., \& {Trevena}, J. 2001, \apj, 556, 121

\bibitem[{{Kewley} {et~al.}(2006){Kewley}, {Groves}, {Kauffmann}, \& {Heckman}}]{Kewley2006AgnClassification}
{Kewley}, L.~J., {Groves}, B., {Kauffmann}, G., \& {Heckman}, T. 2006, \mnras, 372, 961

\bibitem[{{Kewley} {et~al.}(2019){Kewley}, {Nicholls}, \& {Sutherland}}]{Kewley2019EmissionLinesReview}
{Kewley}, L.~J., {Nicholls}, D.~C., \& {Sutherland}, R.~S. 2019, \araa, 57, 511

\bibitem[{{Kikuta} {et~al.}(2019){Kikuta}, {Matsuda}, {Cen}, {Steidel}, {Yagi}, {Hayashino}, {Imanishi}, {Komiyama}, {Momose}, \& {Saito}}]{Kikuta2019LyaViewHyperluminousQso}
{Kikuta}, S., {Matsuda}, Y., {Cen}, R., {et~al.} 2019, \pasj, 71, L2

\bibitem[{{Koposov} {et~al.}(2022){Koposov}, {Speagle}, {Barbary}, {Ashton}, {Bennett}, {Buchner}, {Scheffler}, {Cook}, {Talbot}, {Guillochon}, {Cubillos}, {Asensio Ramos}, {Johnson}, {Lang}, {Ilya}, {Dartiailh}, {Nitz}, {McCluskey}, {Archibald}, {Deil}, {Foreman-Mackey}, {Goldstein}, {Tollerud}, {Leja}, {Kirk}, {Pitkin}, {Sheehan}, {Cargile}, {Ruskin23}, \& {Angus}}]{Koposov2022DynestyZenodoV203}
{Koposov}, S., {Speagle}, J., {Barbary}, K., {et~al.} 2022, {joshspeagle/dynesty: v2.0.3}

\bibitem[{{Koss} {et~al.}(2012){Koss}, {Mushotzky}, {Treister}, {Veilleux}, {Vasudevan}, \& {Trippe}}]{Koss2012UnderstandingDualAgn}
{Koss}, M., {Mushotzky}, R., {Treister}, E., {et~al.} 2012, \apjl, 746, L22

\bibitem[{{Lemaux} {et~al.}(2018){Lemaux}, {Le F{\`e}vre}, {Cucciati}, {Ribeiro}, {Tasca}, {Zamorani}, {Ilbert}, {Thomas}, {Bardelli}, {Cassata}, {Hathi}, {Pforr}, {Smol{\v{c}}i{\'c}}, {Delvecchio}, {Novak}, {Berta}, {McCracken}, {Koekemoer}, {Amor{\'\i}n}, {Garilli}, {Maccagni}, {Schaerer}, \& {Zucca}}]{Lemaux2018VudsProtoCluster}
{Lemaux}, B.~C., {Le F{\`e}vre}, O., {Cucciati}, O., {et~al.} 2018, \aap, 615, A77

\bibitem[{{Leung} {et~al.}(2021){Leung}, {Coil}, {Rupke}, \& {Perrotta}}]{Leung2021KcwMrk273}
{Leung}, G. C.~K., {Coil}, A.~L., {Rupke}, D. S.~N., \& {Perrotta}, S. 2021, \apj, 914, 17

\bibitem[{{Lintott} {et~al.}(2009){Lintott}, {Schawinski}, {Keel}, {van Arkel}, {Bennert}, {Edmondson}, {Thomas}, {Smith}, {Herbert}, {Jarvis}, {Virani}, {Andreescu}, {Bamford}, {Land}, {Murray}, {Nichol}, {Raddick}, {Slosar}, {Szalay}, \& {Vandenberg}}]{Lintott2009HannyVoorwerp}
{Lintott}, C.~J., {Schawinski}, K., {Keel}, W., {et~al.} 2009, \mnras, 399, 129

\bibitem[{{Maiolino} {et~al.}(2023){Maiolino}, {Scholtz}, {Curtis-Lake}, {Carniani}, {Baker}, {de Graaff}, {Tacchella}, {{\"U}bler}, {D'Eugenio}, {Witstok}, {Curti}, {Arribas}, {Bunker}, {Charlot}, {Chevallard}, {Eisenstein}, {Egami}, {Ji}, {Jones}, {Lyu}, {Rawle}, {Robertson}, {Rujopakarn}, {Perna}, {Sun}, {Venturi}, {Williams}, \& {Willott}}]{Maiolino2023JadesPopulationBlackHoles}
{Maiolino}, R., {Scholtz}, J., {Curtis-Lake}, E., {et~al.} 2023, arXiv e-prints, arXiv:2308.01230

\bibitem[{{Mazzarella} {et~al.}(2012){Mazzarella}, {Iwasawa}, {Vavilkin}, {Armus}, {Kim}, {Bothun}, {Evans}, {Spoon}, {Haan}, {Howell}, {Lord}, {Marshall}, {Ishida}, {Xu}, {Petric}, {Sanders}, {Surace}, {Appleton}, {Chan}, {Frayer}, {Inami}, {Khachikian}, {Madore}, {Privon}, {Sturm}, {U}, \& {Veilleux}}]{Mazzarella2012DualAgnMrk266}
{Mazzarella}, J.~M., {Iwasawa}, K., {Vavilkin}, T., {et~al.} 2012, \aj, 144, 125

\bibitem[{{Mazzolari} {et~al.}(2024){Mazzolari}, {{\"U}bler}, {Maiolino}, {Ji}, {Nakajima}, {Feltre}, {Scholtz}, {D'Eugenio}, {Curti}, {Mignoli}, \& {Marconi}}]{Mazzolari24NewAgnDiagnostic}
{Mazzolari}, G., {{\"U}bler}, H., {Maiolino}, R., {et~al.} 2024, arXiv e-prints, arXiv:2404.10811

\bibitem[{{McCarthy} {et~al.}(1987){McCarthy}, {Spinrad}, {Djorgovski}, {Strauss}, {van Breugel}, \& {Liebert}}]{McCarthy1987ExtendedLyaEmissionRadioGal}
{McCarthy}, P.~J., {Spinrad}, H., {Djorgovski}, S., {et~al.} 1987, \apjl, 319, L39

\bibitem[{{Mignoli} {et~al.}(2013){Mignoli}, {Vignali}, {Gilli}, {Comastri}, {Zamorani}, {Bolzonella}, {Bongiorno}, {Lamareille}, {Nair}, {Pozzetti}, {Lilly}, {Carollo}, {Contini}, {Kneib}, {Le F{\`e}vre}, {Mainieri}, {Renzini}, {Scodeggio}, {Bardelli}, {Caputi}, {Cucciati}, {de la Torre}, {de Ravel}, {Franzetti}, {Garilli}, {Iovino}, {Kampczyk}, {Knobel}, {Kova{\v{c}}}, {Le Borgne}, {Le Brun}, {Maier}, {Pell{\`o}}, {Peng}, {Perez Montero}, {Presotto}, {Silverman}, {Tanaka}, {Tasca}, {Tresse}, {Vergani}, {Zucca}, {Bordoloi}, {Cappi}, {Cimatti}, {Koekemoer}, {McCracken}, {Moresco}, \& {Welikala}}]{Mignoli2013ObscuredAgnZcosmos}
{Mignoli}, M., {Vignali}, C., {Gilli}, R., {et~al.} 2013, \aap, 556, A29

\bibitem[{{Miley} \& {De Breuck}(2008)}]{MileyAndDeBreuck2008HzrgsAndEnvironments}
{Miley}, G. \& {De Breuck}, C. 2008, \aapr, 15, 67

\bibitem[{{Miller} {et~al.}(2018){Miller}, {Chapman}, {Aravena}, {Ashby}, {Hayward}, {Vieira}, {Wei{\ss}}, {Babul}, {B{\'e}thermin}, {Bradford}, {Brodwin}, {Carlstrom}, {Chen}, {Cunningham}, {De Breuck}, {Gonzalez}, {Greve}, {Harnett}, {Hezaveh}, {Lacaille}, {Litke}, {Ma}, {Malkan}, {Marrone}, {Morningstar}, {Murphy}, {Narayanan}, {Pass}, {Perry}, {Phadke}, {Rennehan}, {Rotermund}, {Simpson}, {Spilker}, {Sreevani}, {Stark}, {Strandet}, \& {Strom}}]{Miller2018SPT2349}
{Miller}, T.~B., {Chapman}, S.~C., {Aravena}, M., {et~al.} 2018, \nat, 556, 469

\bibitem[{{Monson} {et~al.}(2023){Monson}, {Doore}, {Eufrasio}, {Lehmer}, {Alexander}, {Harrison}, {Kubo}, {Saez}, \& {Umehata}}]{Monson2023XraysSsa22ProtoCluster}
{Monson}, E.~B., {Doore}, K., {Eufrasio}, R.~T., {et~al.} 2023, \apj, 951, 15

\bibitem[{{Nesvadba} {et~al.}(2017){Nesvadba}, {De Breuck}, {Lehnert}, {Best}, \& {Collet}}]{Nesvadba2017SinfoniRadioGalSurvey}
{Nesvadba}, N.~P.~H., {De Breuck}, C., {Lehnert}, M.~D., {Best}, P.~N., \& {Collet}, C. 2017, \aap, 599, A123

\bibitem[{{Nightingale} {et~al.}(2021){Nightingale}, {Hayes}, \& {Griffiths}}]{Nightingale2021PyAutoFit}
{Nightingale}, J., {Hayes}, R., \& {Griffiths}, M. 2021, The Journal of Open Source Software, 6, 2550

\bibitem[{{Noirot} {et~al.}(2018){Noirot}, {Stern}, {Mei}, {Wylezalek}, {Cooke}, {De Breuck}, {Galametz}, {Hatch}, {Vernet}, {Brodwin}, {Eisenhardt}, {Gonzalez}, {Jarvis}, {Rettura}, {Seymour}, \& {Stanford}}]{Noirot2018StructuresCarlaSurvey}
{Noirot}, G., {Stern}, D., {Mei}, S., {et~al.} 2018, \apj, 859, 38

\bibitem[{{Oteo} {et~al.}(2018){Oteo}, {Ivison}, {Dunne}, {Manilla-Robles}, {Maddox}, {Lewis}, {de Zotti}, {Bremer}, {Clements}, {Cooray}, {Dannerbauer}, {Eales}, {Greenslade}, {Omont}, {Perez{\textendash}Fourn{\'o}n}, {Riechers}, {Scott}, {van der Werf}, {Weiss}, \& {Zhang}}]{Oteo2018DRC}
{Oteo}, I., {Ivison}, R.~J., {Dunne}, L., {et~al.} 2018, \apj, 856, 72

\bibitem[{{Overzier}(2016)}]{Overzier2016ProtoclusterReview}
{Overzier}, R.~A. 2016, \aapr, 24, 14

\bibitem[{{Overzier} {et~al.}(2013){Overzier}, {Nesvadba}, {Dijkstra}, {Hatch}, {Lehnert}, {Villar-Mart{\'\i}n}, {Wilman}, \& {Zirm}}]{Overzier2013OpticalEmLinesLyaBlobB1}
{Overzier}, R.~A., {Nesvadba}, N.~P.~H., {Dijkstra}, M., {et~al.} 2013, \apj, 771, 89

\bibitem[{{Pacucci} {et~al.}(2023){Pacucci}, {Nguyen}, {Carniani}, {Maiolino}, \& {Fan}}]{Pacucci2023CeersJadesAgn}
{Pacucci}, F., {Nguyen}, B., {Carniani}, S., {Maiolino}, R., \& {Fan}, X. 2023, \apjl, 957, L3

\bibitem[{{P{\'e}rez-Gonz{\'a}lez} {et~al.}(2024){P{\'e}rez-Gonz{\'a}lez}, {D`Eugenio}, {Rodr{\'\i}guez del Pino}, {{\"U}bler}, {Maiolino}, {Arribas}, {Cresci}, {Lamperti}, {Bunker}, {Carniani}, {Charlot}, {Willott}, {B{\"o}ker}, {Parlanti}, {Scholtz}, {Venturi}, {Barro}, {Costantin}, {Mart{\'\i}n-Navarro}, {Dunlop}, \& {Magee}}]{PerezGonzalez2024NirspecJekyllHyde}
{P{\'e}rez-Gonz{\'a}lez}, P.~G., {D`Eugenio}, F., {Rodr{\'\i}guez del Pino}, B., {et~al.} 2024, arXiv e-prints, arXiv:2405.03744

\bibitem[{{Perna} {et~al.}(2023){Perna}, {Arribas}, {Marshall}, {D'Eugenio}, {{\"U}bler}, {Bunker}, {Charlot}, {Carniani}, {Jakobsen}, {Maiolino}, {Rodr{\'\i}guez Del Pino}, {Willott}, {B{\"o}ker}, {Circosta}, {Cresci}, {Curti}, {Husemann}, {Kumari}, {Lamperti}, {P{\'e}rez-Gonz{\'a}lez}, \& {Scholtz}}]{Perna2023GaNifsDualAgnEnvironment}
{Perna}, M., {Arribas}, S., {Marshall}, M., {et~al.} 2023, \aap, 679, A89

\bibitem[{{Reines} \& {Volonteri}(2015)}]{ReinesAndVolonteri2015BhMassVsMstar}
{Reines}, A.~E. \& {Volonteri}, M. 2015, \apj, 813, 82

\bibitem[{{Reuland} {et~al.}(2003){Reuland}, {van Breugel}, {R{\"o}ttgering}, {de Vries}, {Stanford}, {Dey}, {Lacy}, {Bland-Hawthorn}, {Dopita}, \& {Miley}}]{Reuland2003GiantLyaNebulaeRadioGal}
{Reuland}, M., {van Breugel}, W., {R{\"o}ttgering}, H., {et~al.} 2003, \apj, 592, 755

\bibitem[{{Rich} {et~al.}(2011){Rich}, {Kewley}, \& {Dopita}}]{Rich2011GalaxyWideShocksLirgs}
{Rich}, J.~A., {Kewley}, L.~J., \& {Dopita}, M.~A. 2011, \apj, 734, 87

\bibitem[{{Rich} {et~al.}(2015){Rich}, {Kewley}, \& {Dopita}}]{Rich2015MergersDriveShocksGoals}
{Rich}, J.~A., {Kewley}, L.~J., \& {Dopita}, M.~A. 2015, \apjs, 221, 28

\bibitem[{{Riechers} {et~al.}(2014){Riechers}, {Carilli}, {Capak}, {Scoville}, {Smol{\v{c}}i{\'c}}, {Schinnerer}, {Yun}, {Cox}, {Bertoldi}, {Karim}, \& {Yan}}]{Riechers2014AlmaCiiAztec3}
{Riechers}, D.~A., {Carilli}, C.~L., {Capak}, P.~L., {et~al.} 2014, \apj, 796, 84

\bibitem[{{Rigby} {et~al.}(2023){Rigby}, {Vieira}, {Phadke}, {Hutchison}, {Welch}, {Cathey}, {Spilker}, {Gonzalez}, {Adhikari}, {Aravena}, {Bayliss}, {Birkin}, {Bursk}, {Chapman}, {Dahle}, {Elicker}, {Fischer}, {Florian}, {Gladders}, {Hayward}, {Hewald}, {Kettler}, {Khullar}, {Kim}, {Law}, {Mahler}, {Malhotra}, {Murphy}, {Narayanan}, {Olivier}, {Rhoads}, {Sharon}, {Solimano}, {Thiruvengadam}, {Vizgan}, \& {Younker}}]{Rigby2023TemplatesOverview}
{Rigby}, J.~R., {Vieira}, J.~D., {Phadke}, K.~A., {et~al.} 2023, arXiv e-prints, arXiv:2312.10465

\bibitem[{{Roy} {et~al.}(2024){Roy}, {Heckman}, {Overzier}, {Saxena}, {Duncan}, {Miley}, {Villar Mart{\'\i}n}, {Gab{\'a}nyi}, {Aydar}, {Bosman}, {Rottgering}, {Pentericci}, {Onoue}, \& {Reynaldi}}]{Roy2024JwstFeedbackRadioJets}
{Roy}, N., {Heckman}, T., {Overzier}, R., {et~al.} 2024, \apj, 970, 69

\bibitem[{{Saxena} {et~al.}(2024){Saxena}, {Overzier}, {Villar-Mart{\'\i}n}, {Heckman}, {Roy}, {Duncan}, {R{\"o}ttgering}, {Miley}, {Aydar}, {Best}, {Bosman}, {Cameron}, {Gab{\'a}nyi}, {Humphrey}, {Morais}, {Onoue}, {Pentericci}, {Reynaldi}, \& {Venemans}}]{Saxena2024WidespreadFeedbackProtoBcgJwst}
{Saxena}, A., {Overzier}, R.~A., {Villar-Mart{\'\i}n}, M., {et~al.} 2024, \mnras, 531, 4391

\bibitem[{{Schaerer} {et~al.}(2019){Schaerer}, {Fragos}, \& {Izotov}}]{Schaerer2019XrayBinariesOriginHeII}
{Schaerer}, D., {Fragos}, T., \& {Izotov}, Y.~I. 2019, \aap, 622, L10

\bibitem[{{Schawinski} {et~al.}(2007){Schawinski}, {Thomas}, {Sarzi}, {Maraston}, {Kaviraj}, {Joo}, {Yi}, \& {Silk}}]{Schawinski2007AgnFeedbackSdssLiners}
{Schawinski}, K., {Thomas}, D., {Sarzi}, M., {et~al.} 2007, \mnras, 382, 1415

\bibitem[{{Scholtz} {et~al.}(2023){Scholtz}, {Maiolino}, {D'Eugenio}, {Curtis-Lake}, {Carniani}, {Charlot}, {Curti}, {Silcock}, {Arribas}, {Baker}, {Bhatawdekar}, {Boyett}, {Bunker}, {Chevallard}, {Circosta}, {Eisenstein}, {Hainline}, {Hausen}, {Ji}, {Ji}, {Johnson}, {Kumari}, {Looser}, {Lyu}, {Maseda}, {Parlanti}, {Perna}, {Rieke}, {Robertson}, {Rodr{\'\i}guez Del Pino}, {Sun}, {Tacchella}, {{\"U}bler}, {Venturi}, {Williams}, {Willmer}, {Willott}, \& {Witstok}}]{Scholtz2023JadesLargeAgnPop}
{Scholtz}, J., {Maiolino}, R., {D'Eugenio}, F., {et~al.} 2023, arXiv e-prints, arXiv:2311.18731

\bibitem[{{Schwarz}(1978)}]{Schwarz1978BayesianInfoCriterion}
{Schwarz}, G. 1978, Annals of Statistics, 6, 461

\bibitem[{{Secrest} {et~al.}(2017){Secrest}, {Schmitt}, {Blecha}, {Rothberg}, \& {Fischer}}]{Secrest2017Was49bLocalOffsetAgn}
{Secrest}, N.~J., {Schmitt}, H.~R., {Blecha}, L., {Rothberg}, B., \& {Fischer}, J. 2017, \apj, 836, 183

\bibitem[{{Shen} {et~al.}(2007){Shen}, {Strauss}, {Oguri}, {Hennawi}, {Fan}, {Richards}, {Hall}, {Gunn}, {Schneider}, {Szalay}, {Thakar}, {Vanden Berk}, {Anderson}, {Bahcall}, {Connolly}, \& {Knapp}}]{Shen2007SdssQsoClustering}
{Shen}, Y., {Strauss}, M.~A., {Oguri}, M., {et~al.} 2007, \aj, 133, 2222

\bibitem[{{Shirazi} \& {Brinchmann}(2012)}]{ShiraziAndBrinchmann2012SdssHeIIemission}
{Shirazi}, M. \& {Brinchmann}, J. 2012, \mnras, 421, 1043

\bibitem[{{Smith} {et~al.}(2022){Smith}, {Krause}, {Hardcastle}, \& {Drake}}]{Smith2022RadioRelicHsVLofar}
{Smith}, D.~J.~B., {Krause}, M.~G., {Hardcastle}, M.~J., \& {Drake}, A.~B. 2022, \mnras, 514, 3879

\bibitem[{{Smol{\v{c}}i{\'c}} {et~al.}(2015){Smol{\v{c}}i{\'c}}, {Karim}, {Miettinen}, {Novak}, {Magnelli}, {Riechers}, {Schinnerer}, {Capak}, {Bondi}, {Ciliegi}, {Aravena}, {Bertoldi}, {Bourke}, {Banfield}, {Carilli}, {Civano}, {Ilbert}, {Intema}, {Le F{\`e}vre}, {Finoguenov}, {Hallinan}, {Kl{\"o}ckner}, {Koekemoer}, {Laigle}, {Masters}, {McCracken}, {Mooley}, {Murphy}, {Navarette}, {Salvato}, {Sargent}, {Sheth}, {Toft}, \& {Zamorani}}]{Smolcic2015PhysicalPropertiesSmgsCosmos}
{Smol{\v{c}}i{\'c}}, V., {Karim}, A., {Miettinen}, O., {et~al.} 2015, \aap, 576, A127

\bibitem[{{Smol{\v{c}}i{\'c}} {et~al.}(2017{\natexlab{a}}){Smol{\v{c}}i{\'c}}, {Miettinen}, {Tomi{\v{c}}i{\'c}}, {Zamorani}, {Finoguenov}, {Lemaux}, {Aravena}, {Capak}, {Chiang}, {Civano}, {Delvecchio}, {Ilbert}, {Jurlin}, {Karim}, {Laigle}, {Le F{\`e}vre}, {Marchesi}, {McCracken}, {Riechers}, {Salvato}, {Schinnerer}, {Tasca}, \& {Toft}}]{Smolcic2017SmgEnvironments}
{Smol{\v{c}}i{\'c}}, V., {Miettinen}, O., {Tomi{\v{c}}i{\'c}}, N., {et~al.} 2017{\natexlab{a}}, \aap, 597, A4

\bibitem[{{Smol{\v{c}}i{\'c}} {et~al.}(2017{\natexlab{b}}){Smol{\v{c}}i{\'c}}, {Novak}, {Delvecchio}, {Ceraj}, {Bondi}, {Delhaize}, {Marchesi}, {Murphy}, {Schinnerer}, {Vardoulaki}, \& {Zamorani}}]{Smolcic2017VlaCosmosEvolutionOfRadioAgn}
{Smol{\v{c}}i{\'c}}, V., {Novak}, M., {Delvecchio}, I., {et~al.} 2017{\natexlab{b}}, \aap, 602, A6

\bibitem[{{Solimano} {et~al.}(2024){Solimano}, {Gonz{\'a}lez-L{\'o}pez}, {Aravena}, {Herrera-Camus}, {De Looze}, {F{\"o}rster Schreiber}, {Spilker}, {Tadaki}, {Assef}, {Barcos-Mu{\~n}oz}, {Davies}, {D{\'\i}az-Santos}, {Ferrara}, {Fisher}, {Guaita}, {Ikeda}, {Johnston}, {Lutz}, {Mitsuhashi}, {Moya-Sierralta}, {Rela{\~n}o}, {Naab}, {Posses}, {Telikova}, {{\"U}bler}, {van der Giessen}, {Veilleux}, \& {Villanueva}}]{Solimano2024CristalPlume}
{Solimano}, M., {Gonz{\'a}lez-L{\'o}pez}, J., {Aravena}, M., {et~al.} 2024, \aap, 689, A145

\bibitem[{{Speagle}(2020)}]{Speagle2020Dynesty}
{Speagle}, J.~S. 2020, \mnras, 493, 3132

\bibitem[{{Staab} {et~al.}(2024){Staab}, {Lemaux}, {Forrest}, {Shah}, {Cucciati}, {Lubin}, {Gal}, {Hung}, {Shen}, {Giddings}, {Khusanova}, {Zamorani}, {Bardelli}, {Cassara}, {Cassata}, {Chiang}, {Fudamoto}, {Fukushima}, {Garilli}, {Giavalisco}, {Gruppioni}, {Guaita}, {Gururajan}, {Hathi}, {Kashino}, {Scoville}, {Talia}, {Vergani}, \& {Zucca}}]{Staab2024TaralayProtocluster}
{Staab}, P., {Lemaux}, B.~C., {Forrest}, B., {et~al.} 2024, \mnras, 528, 6934

\bibitem[{{Stockton} {et~al.}(2006){Stockton}, {Fu}, \& {Canalizo}}]{Stockton2006QsoExtendedEmissionLineRegions}
{Stockton}, A., {Fu}, H., \& {Canalizo}, G. 2006, \nar, 50, 694

\bibitem[{{Storey} \& {Zeippen}(2000)}]{StoreyAndZeippen2000TheoreticalOiiiDoubletRatio}
{Storey}, P.~J. \& {Zeippen}, C.~J. 2000, \mnras, 312, 813

\bibitem[{{Swinbank} {et~al.}(2015){Swinbank}, {Vernet}, {Smail}, {De Breuck}, {Bacon}, {Contini}, {Richard}, {R{\"o}ttgering}, {Urrutia}, \& {Venemans}}]{Swinbank2015MappingGiantLyaHaloRadioGal}
{Swinbank}, A.~M., {Vernet}, J.~D.~R., {Smail}, I., {et~al.} 2015, \mnras, 449, 1298

\bibitem[{{{\"U}bler} {et~al.}(2024){{\"U}bler}, {Maiolino}, {P{\'e}rez-Gonz{\'a}lez}, {D'Eugenio}, {Perna}, {Curti}, {Arribas}, {Bunker}, {Carniani}, {Charlot}, {Rodr{\'\i}guez Del Pino}, {Baker}, {B{\"o}ker}, {Cresci}, {Dunlop}, {Grogin}, {Jones}, {Kumari}, {Lamperti}, {Laporte}, {Marshall}, {Mazzolari}, {Parlanti}, {Rawle}, {Scholtz}, {Venturi}, \& {Witstok}}]{Ubler2024OffsetAgnGaNifs}
{{\"U}bler}, H., {Maiolino}, R., {P{\'e}rez-Gonz{\'a}lez}, P.~G., {et~al.} 2024, \mnras, 531, 355

\bibitem[{{Umehata} {et~al.}(2021){Umehata}, {Smail}, {Steidel}, {Hayes}, {Scott}, {Swinbank}, {Ivison}, {Nagao}, {Kubo}, {Nakanishi}, {Matsuda}, {Ikarashi}, {Tamura}, \& {Geach}}]{Umehata2021CiiObsOfLAB1}
{Umehata}, H., {Smail}, I., {Steidel}, C.~C., {et~al.} 2021, \apj, 918, 69

\bibitem[{{Veilleux} \& {Osterbrock}(1987)}]{VeilleuxAndOsterbrock1987SpectralClassGalaxies}
{Veilleux}, S. \& {Osterbrock}, D.~E. 1987, \apjs, 63, 295

\bibitem[{{Venemans} {et~al.}(2007){Venemans}, {R{\"o}ttgering}, {Miley}, {van Breugel}, {de Breuck}, {Kurk}, {Pentericci}, {Stanford}, {Overzier}, {Croft}, \& {Ford}}]{Venemans2007ProtoclustersHighzRadioGals}
{Venemans}, B.~P., {R{\"o}ttgering}, H.~J.~A., {Miley}, G.~K., {et~al.} 2007, \aap, 461, 823

\bibitem[{{Vito} {et~al.}(2024){Vito}, {Brandt}, {Comastri}, {Gilli}, {Ivison}, {Lanzuisi}, {Lehmer}, {Lopez}, {Tozzi}, \& {Vignali}}]{Vito2024XraySmbhSpt2349}
{Vito}, F., {Brandt}, W.~N., {Comastri}, A., {et~al.} 2024, \aap, 689, A130

\bibitem[{{Vito} {et~al.}(2020){Vito}, {Brandt}, {Lehmer}, {Vignali}, {Zou}, {Bauer}, {Bremer}, {Gilli}, {Ivison}, \& {Spingola}}]{Vito2020ComptonThickQsoInDrc}
{Vito}, F., {Brandt}, W.~N., {Lehmer}, B.~D., {et~al.} 2020, \aap, 642, A149

\bibitem[{{Wang} {et~al.}(2021){Wang}, {Hill}, {Chapman}, {Wei{\ss}}, {Scott}, {Apostolovski}, {Aravena}, {Archipley}, {B{\'e}thermin}, {Canning}, {De Breuck}, {Dong}, {Everett}, {Gonzalez}, {Greve}, {Hayward}, {Hezaveh}, {Jarugula}, {Marrone}, {Phadke}, {Reuter}, {Rotermund}, {Spilker}, \& {Vieira}}]{Wang2021OverdensitiesSubmmSpt}
{Wang}, G. C.~P., {Hill}, R., {Chapman}, S.~C., {et~al.} 2021, \mnras, 508, 3754

\bibitem[{{Wang} {et~al.}(2024){Wang}, {Wylezalek}, {De Breuck}, {Vernet}, {Rupke}, {Zakamska}, {Vayner}, {Lehnert}, {Nesvadba}, \& {Stern}}]{Wang2024AgnIonizationConeRadioLoud}
{Wang}, W., {Wylezalek}, D., {De Breuck}, C., {et~al.} 2024, \aap, 683, A169

\bibitem[{{Weaver} {et~al.}(2022){Weaver}, {Kauffmann}, {Ilbert}, {McCracken}, {Moneti}, {Toft}, {Brammer}, {Shuntov}, {Davidzon}, {Hsieh}, {Laigle}, {Anastasiou}, {Jespersen}, {Vinther}, {Capak}, {Casey}, {McPartland}, {Milvang-Jensen}, {Mobasher}, {Sanders}, {Zalesky}, {Arnouts}, {Aussel}, {Dunlop}, {Faisst}, {Franx}, {Furtak}, {Fynbo}, {Gould}, {Greve}, {Gwyn}, {Kartaltepe}, {Kashino}, {Koekemoer}, {Kokorev}, {Le F{\`e}vre}, {Lilly}, {Masters}, {Magdis}, {Mehta}, {Peng}, {Riechers}, {Salvato}, {Sawicki}, {Scarlata}, {Scoville}, {Shirley}, {Silverman}, {Sneppen}, {Smolc̆i{\'c}}, {Steinhardt}, {Stern}, {Tanaka}, {Taniguchi}, {Teplitz}, {Vaccari}, {Wang}, \& {Zamorani}}]{Weaver2022Cosmos2020}
{Weaver}, J.~R., {Kauffmann}, O.~B., {Ilbert}, O., {et~al.} 2022, \apjs, 258, 11

\bibitem[{{Wylezalek} {et~al.}(2013){Wylezalek}, {Galametz}, {Stern}, {Vernet}, {De Breuck}, {Seymour}, {Brodwin}, {Eisenhardt}, {Gonzalez}, {Hatch}, {Jarvis}, {Rettura}, {Stanford}, \& {Stevens}}]{Wylezalek2013SpitzerRadioLoudAgn}
{Wylezalek}, D., {Galametz}, A., {Stern}, D., {et~al.} 2013, \apj, 769, 79

\bibitem[{{Wylezalek} {et~al.}(2022){Wylezalek}, {Vayner}, {Rupke}, {Zakamska}, {Veilleux}, {Ishikawa}, {Bertemes}, {Liu}, {Barrera-Ballesteros}, {Chen}, {Goulding}, {Greene}, {Hainline}, {Hamann}, {Heckman}, {Johnson}, {Lutz}, {L{\"u}tzgendorf}, {Mainieri}, {Maiolino}, {Nesvadba}, {Ogle}, \& {Sturm}}]{Wylezalek2022NirspecExtremelyRedQso}
{Wylezalek}, D., {Vayner}, A., {Rupke}, D. S.~N., {et~al.} 2022, \apjl, 940, L7

\end{thebibliography}

\begin{appendix}
\section{Line fitting with pPXF}\label{sec:ap:ppxf}
We use the template-fitting software \verb|pPXF| \citep[version 9.2.1;][]{Cappellari2017FullSpectrumFitting, Cappellari2023PpxfUpdate} to simultaneously model the continuum and the emission lines of our aperture-extracted spectra in the full wavelength range covered by a single grating.
We start with the G395H grating since it provides better spectral resolution than G235M.
For each spectrum, we perform perform fits with one and two velocity components for the emission lines, but a single component for the stars, which is tied to the narrow gas component. We include the following lines in the fit:
\permitted{He}{i}{5877}, \doublet{O}{i}{6303}{6365}, \doublet{N}{ii}{6550}{6585}, \permitted{H}{i}{6565} (\halpha), \doublet{S}{ii}{6718}{6732}, \permitted{He}{i}{7065}, \singlet{Ar}{iii}{7138}, and \singlet{S}{iii}{9071}, where the [O\,{\sc i}] and [N\,{\sc ii}] doublet ratios have been fixed to their theoretical values.
The continuum is fitted against a grid of Stellar Population Synthesis (SPS) spectra computed with \verb|fsps| v3.2 \citep{Conroy2009FspsImf, Conroy2010FspsCalib}, but restricted to ages younger than the age of the Universe at $z=4.54$.
In most cases, the continuum has S/N$\lesssim 1$ per resolution element and no stellar absorption features can be identified, hence we refrain from interpreting any of the SPS output parameters.
We find, nevertheless, that these templates provide a good representation of the continuum slope, and naturally incorporate the stellar absorption correction for the Balmer emission lines, even though this correction always stays below 1\%. %double check!!!

We then compute the AIC and BIC scores of both single and double component fits, and require the score difference to be larger than five to keep the double component fit as the preferred model. This criterion is only met for O3-N and O3-N-core.

Next, we model the G235M spectrum using the velocity and velocity dispersion best fit values from the G395H fit as starting values.
Here, we fit the continuum with the same libraries as above, and include the following list of emission lines:
\singlet{Ne}{v}{3427}, \doublet{O}{ii}{3727}{3730}, \singlet{Ne}{iii}{3870}, \singlet{Ne}{iii}{3969}, \singlet{O}{iii}{4364}, \permitted{He}{ii}{4687}, \doublet{O}{iii}{4960}{5008}, and the Balmer series from \permitted{H}{i}{3799} (H10) to \permitted{H}{i}{4863} (\hbeta).
The relative intensities of the \oiii doublet are fixed to their theoretical ratio.

Throughout the paper, we use the line fluxes and uncertainties measured by \verb|pPXF| to compute line ratios. The values are presented in Table \ref{tab:line_fluxes} and represent model fluxes from the single Gaussian component fits, except for O3-N and O3-N-core, where we use the sum of the narrow and broad components. This is because in most lines of the broad component have too low a S/N to provide a meaningful ratio on their own. We also quote fluxes and ratios without correction by reddening unless otherwise noted.

%\section{Adaptive moment maps}\label{sec:ap:moments}
%The velocity field and velocity dispersion maps shown in Fig.~\ref{fig:o3-moments} were created following \citet{Solimano2024CristalPlume}, with a spatial and spectral Gaussian convolution kernel applied to the continuum subtracted cube. The spatial kernel has $\sigma=1$ spaxel, whereas the spectral kernel has $\sigma=\sigma_{LSF}$ at the wavelength of the line. The moments are created by masking out all the voxels with S/N$<3$ in the convolved cube.

\section{Double Gaussian fitting}\label{sec:ap:gaussfit}
In this appendix we describe the method used to fit the line profiles presented in Sec.~\ref{sec:results:broad}. In the case of \oiii, we model both lines in the doublet simultaneously but tie their wavelengths and amplitudes to the expected ratios (e.g., \singlet{O}{iii}{5008}/\singlet{O}{iii}{4960}$=2.98$, \citealt{StoreyAndZeippen2000TheoreticalOiiiDoubletRatio}). For \halpha, we also fit the \nii doublet with the \singlet{N}{ii}{6585}/\singlet{N}{ii}{6550} ratio fixed to \num{2.8}.
%Moreover, we set the two lines to share the same velocity width. Thus, our single-component model has four free parameters, while the double-component model has seven free parameters.
%We also set broad uniform priors on each parameter based on the data ranges, except for the line center, where we use a Gaussian prior centered at zero velocity and standard deviation of \SI{500}{\kilo\meter\per\second} for quicker convergence.
We set up the models within the probabilistic programming framework \textsc{PyAutoFit}  \citep[version 2024.1.27.4,][]{Nightingale2021PyAutoFit}, and use the \textsc{Dynesty} \citep{Speagle2020Dynesty, Koposov2022DynestyZenodoV203} backend to sample the posterior probability distribution and estimate the Bayesian evidence $\log\left (Z\right)$.
The width of the line spread function (LSF) is taken from the dispersion curves available in the JWST documentation\footnote{\href{https://jwst-docs.stsci.edu/jwst-near-infrared-spectrograph/nirspec-instrumentation/nirspec-dispersers-and-filters\#NIRSpecDispersersandFilters-DispersioncurvesfortheNIRSpecdispersers\&gsc.tab=0}{NIRSpec Dispersers and Filters}}.

\section{Faint lines}\label{sec:ap:faint}
\begin{figure}[!htb]
    \centering
    \resizebox{\hsize}{!}{\includegraphics{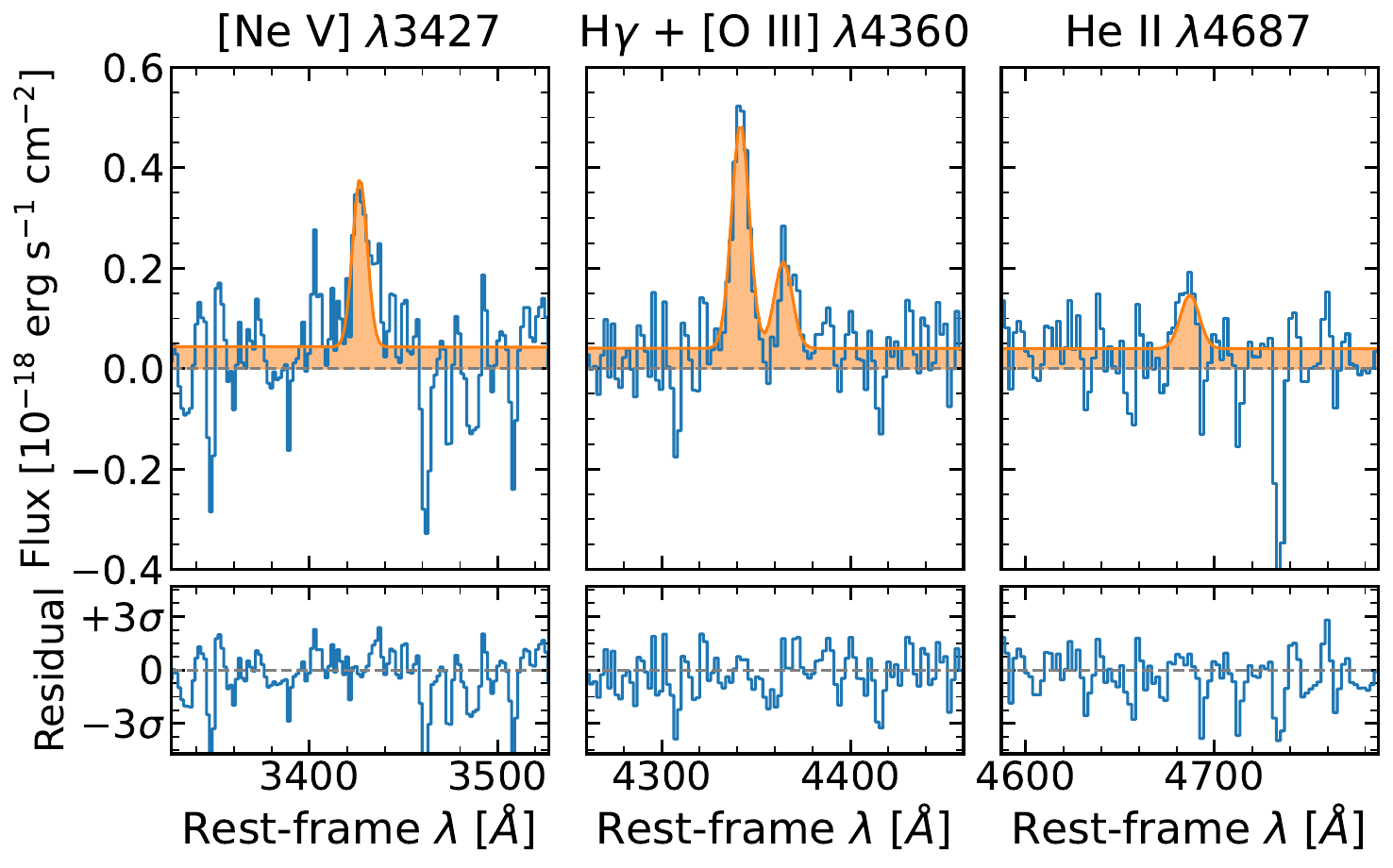}}
    \caption{Zoom-in to high-ionization lines detected in O3-N-core. The orange filled curve denotes the pPXF best fit stellar plus gaseous template.}
    \label{fig:nev-O3-N-core}
\end{figure}

\begin{figure}[!hbt]
    \centering
    \resizebox{\hsize}{!}{\includegraphics{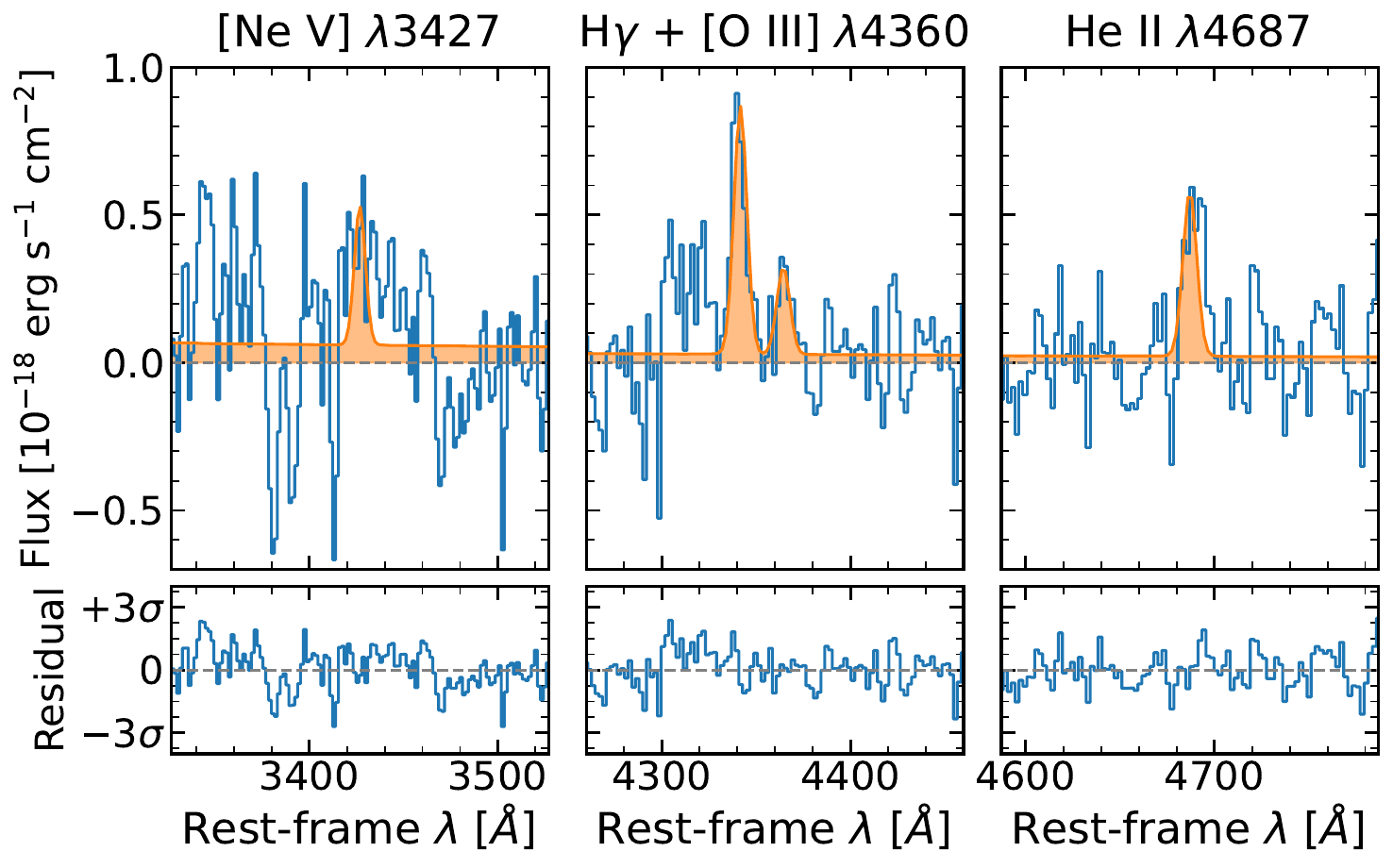}}
    \caption{Same as Fig.~\ref{fig:nev-O3-N-core} but for O3-S-HeII. Here, the detection of the [Ne\, \textsc{v}] line is only tentative.}
    \label{fig:nev-O3-S-HeII}
\end{figure}

In this appendix we plot spectral cutouts of O3-N-core (Fig.~\ref{fig:nev-O3-N-core}) and O3-S-HeII (Fig.~\ref{fig:nev-O3-S-HeII}) around two high-ionization lines (\heii and [Ne\,{\sc ii}]) plus the \singlet{O}{iii}{4364} auroral line.

\begin{table*}[!htb]
    \centering
    \caption{Line fluxes for the different apertures used in this paper.}
    \resizebox{17cm}{!}{%\input{tables/line_fluxes}
	\begin{tabular}{lcccccccc}
	\hline
	Aperture & O3-N & O3-N-core & O3-S & O3-S-HeII & C01-total & C01-NE & C01-SW & DSFG \\
	\hline
	R. A. (deg) & \num{150.227098} & \num{150.227084} & \num{150.226878} & \num{150.226861} & \num{150.227185} & \num{150.227227} & \num{150.227145} & \num{150.227038} \\
	Dec. (deg) &\num{2.576809} & \num{2.576772} & \num{2.576298} & \num{2.576304} & \num{2.576316} & \num{2.576378} & \num{2.576221} & \num{2.576655} \\
	\oiii redshift &\num[separate-uncertainty=false]{4.5433+-0.0001} & \num[separate-uncertainty=false]{4.5418+-0.0001} & \num[separate-uncertainty=false]{4.5448+-0.0003} & \num[separate-uncertainty=false]{4.5516+-0.0001} & \num[separate-uncertainty=false]{4.5527+-0.0002} & \num[separate-uncertainty=false]{4.5521+-0.0001} & \num[separate-uncertainty=false]{4.5414+-0.0001} & \num[separate-uncertainty=false]{4.5445+-0.0001} \\
	\hline
	\singlet{Ne}{v}{3427} & $<15.4$ & $2.3\pm0.3$ & $6\pm2$ & $<2.2$ & $<3.7$ & $<2.3$ & $<1.5$ & $<2.7$ \\
	\doublet{O}{ii}{3727}{30} & $35\pm2$\tablefootmark{a} & $6\pm2$ & $35\pm9$ & $13\pm3$ & $30\pm6$ & $11.3\pm1.4$\tablefootmark{a} & $13\pm2$ & $7.0\pm0.7$\tablefootmark{a} \\
	\singlet{Ne}{iii}{3870} & $19\pm3.4$ & $4.6\pm0.2$ & $15.9\pm1.2$ & $5.1\pm 0.5$ & $6.7 \pm 1.0$ & $<2.0$  & $3.3\pm0.4$ & $2.4\pm 0.6$ \\
	H$\gamma\,\lambda4342$ & $13\pm3$ & $2.6\pm0.2$ & $8\pm1$ & $3.7\pm0.4$ & $4.9\pm0.9$ & $2.2\pm0.5$ & $2.0\pm0.3$ & $1.9\pm0.5$ \\
	\singlet{O}{iii}{4364} & $<8.5$ & $1.0\pm0.2$ & $<3.1$ & $1.3\pm0.4$ & $<2.6$ & $<1.7$ & $<1.0$ & $<1.6$ \\
	\permitted{He}{ii}{4687} & $<8.4$ & $0.6\pm0.2$ & $5\pm1$ & $2.4\pm0.5$ & $<2.5$ & $<1.7$ & $<0.9$ & $1.9\pm0.5$ \\
	H$\beta\,\lambda4863$ & $25\pm3$ & $6.0\pm0.2$ & $25\pm1$ & $8.5\pm0.5$ & $12.0\pm0.9$ & $4.6\pm0.6$ & $5.2\pm0.3$ & $3.5\pm0.5$ \\
	\singlet{O}{iii}{5008} & $251\pm4$ & $60.0\pm0.5$ & $223\pm2$ & $87.4\pm0.9$ & $79\pm1$ & $29.0\pm0.8$ & $35.1\pm0.5$ & $31.8\pm0.6$ \\
	\doublet{O}{i}{6302}{66} & $6.4\pm0.6$ & $1.22\pm0.05$ & $4.2\pm0.4$ & $0.97\pm0.09$ & $<0.6$ & $<0.3$ & $0.76\pm0.09$ & $1.3\pm0.1$ \\
	H$\alpha\,\lambda6565$ & $122.5\pm0.8$ & $26.9\pm0.1$ & $78.5\pm0.4$ & $23.0\pm0.1$ & $45.1\pm0.2$ & $18.2\pm0.1$ & $18.4\pm0.1$ & $17.4\pm0.1$ \\
	\doublet{N}{ii}{6550}{85} & $15.0\pm0.8$ & $5.0\pm0.1$ & $2.4\pm0.4$ & $0.9\pm0.1$ & $4.6\pm0.2$ & $2.2\pm0.1$ & $0.78\pm0.09$ & $20.9\pm0.2$ \\
	\doublet{S}{ii}{6718}{33} & $14\pm3$ & $2.3\pm0.3$ & $11\pm1$ & $2.5\pm0.3$ & $5.9\pm0.7$ & $1.8\pm0.3$ & $2.4\pm0.2$ & $5.2\pm0.6$ \\
	\hline
	\hline
	\end{tabular}
    }
    \tablefoot{All fluxes are given in units of \SI{e-18}{\erg\per\second\per\centi\meter\squared} and are not corrected for dust reddening. Upper limits are quoted at the $3\sigma$ level.\\
    \tablefoottext{a}{Measured in standalone single-component fit to avoid unrealistically large uncertainty.}}
    \label{tab:line_fluxes}
\end{table*}
\end{appendix}

\end{document}